# Supercurrent effect in a charge density wave intertwined superconductor

Zhen Zhu[1]†, Wei Cheng[2]†, Dang Liu[1]†, Pengyu Hu[1], Yi Yang[1], Qiaoyan Yu[1], Shasha Xue[1], Ruijun Xi[1], Xingsen Chen[1], Jice Sun[1], Dandan Guan[1,4,5], Yaoyi Li[1,4,5], Shiyong Wang[1,4,5], Canhua Liu[1,4,5], Zhuan Xu[3], Xin Liu[1,2,4]*, Hao Zheng[1,4,5]* & Jinfeng Jia[1,4,5,6,7]*

1. State Key Laboratory of Micro-nano Engineering Science, Tsung-Dao Lee Institute & School of Physics and Astronomy, Key Laboratory of Artificial Structures and Quantum Control (Ministry of Education), Shanghai Jiao Tong University, Shanghai 200240, China
2. School of Physics and Wuhan National High Magnetic Field Center, Huazhong University of Science and Technology, Wuhan, China
3. Department of Physics, Zhejiang University, Hangzhou, Zhejiang, China
4. Hefei National Laboratory, Hefei 230088, China
5. Shanghai Research Center for Quantum Sciences, 99 Xiupu Road, Shanghai 201315, China
6. Department of Physics, Southern University of Science and Technology, Shenzhen, China
7. Quantum Science Center of Guangdong-Hong Kong-Macao Greater Bay Area (Guangdong), Shenzhen, China

†These authors contributed equally to this work.

*Corresponding authors: haozheng1@sjtu.edu.cn ;phyliuxin@sjtu.edu.cn ; jfjia@sjtu.edu.cn

**Abstract**

The energy-momentum ($E$-k) dispersion of quasiparticles constitutes a fundamental concept in condensed matter systems. The ability to modify the $E$-k dispersion, exemplified by supercurrent-induced Doppler shifts of Bogoliubov quasiparticle spectra in superconductors, enables manipulation of various emergent quantum properties. However, investigations into the supercurrent effect on superconductors intertwined with charge orders remain scarce. Here, we report that the Meissner current, generated by the diamagnetic response to an applied in-plane magnetic field, can tailor Bogoliubov quasiparticle excitations at the precursor charge density wave (CDW) vectors. Our scanning tunneling spectroscopic imaging reveals a field-driven symmetry breaking of CDW modulations, specifically a $C_{3v}$-to-$C_s$ transition, in superconducting $NbSe_2$. Model calculations suggest that the observed anisotropy originates from a selective Doppler-shift-induced $E$-k dispersion reconstruction. Furthermore, altering the field direction enables on-demand tuning of anisotropic CDW modulations and visualization of their momentum-space distribution. These results highlight a novel mechanism for controlling emergent electronic phases through momentum-space engineering.

## Introduction

Bogoliubov quasiparticles are the elementary excitations in superconductors and obey the energy-momentum dispersion: $E(\mathbf{k}) = \sqrt{(\epsilon_{\mathbf{k}} - \mu)^2 + \Delta^2}$, where $\epsilon_{\mathbf{k}}$ denotes the normal-state dispersion, $\mu$ is the chemical potential and $\Delta$ is the superconducting gap. When a supercurrent flows through a superconductor, the dispersion of Bogoliubov quasiparticles acquires an additional term[1], i.e., the Doppler shift $E_{\mathrm{D}} = e\mathbf{v}_{\mathbf{k}} \cdot \mathbf{A}$ ($\mathbf{v}_{\mathbf{k}}$ is the Fermi velocity at momentum $\mathbf{k}$, $\mathbf{A}$ is the vector potential), which gives rise to many novel phenomena, including segmented Fermi surfaces[2], the Josephson diode effect[3], and topological quantum phase transitions[4,5]. Within engineered structures, supercurrents may also be harnessed to create and manipulate Majorana zero modes[6-9]. Furthermore, the Doppler shift can tailor the optical and thermal properties of superconductors[10-12]. Like superconductivity, CDW represents another prototypical symmetry-broken phase, arising from Fermi surface instabilities[13]. CDW and superconductivity can either compete or coexist in quantum materials, and their interplay has captivated research interest for decades[14-19]. In CDW intertwined superconductors, CDW order may superimpose a spatial modulation on the Bogoliubov quasiparticles in real space, thereby breaking the underlying crystal translational symmetry. However, the effect of supercurrents in such superconductors remains poorly understood. Here, we employ in-plane magnetic fields to generate shielding currents at the sample surface through the Meissner effect, and utilize dilution refrigerator–based scanning tunneling microscopy and spectroscopy (STM/STS) to reveal the supercurrent-induced symmetry breaking in momentum space and its impact on the orientation- and energy-dependent characteristics of Bogoliubov quasiparticle excitations at the precursor CDW vectors in 2H-$NbSe_2$.

2H-$NbSe_2$ is a layered transition-metal dichalcogenide comprises of hexagonally coordinated Se–Nb–Se trilayers stacked along the *c*-axis by weak van der Waals forces. It exhibits s-wave superconductivity with $T_c \approx 7.2$ K and undergoes a 3×3 CDW phase transition at $T_{\mathrm{CDW}} \approx 33$ K. The coexistence of superconductivity and CDW order has established $NbSe_2$ as a canonical system in condensed matter physics[20-22], underpinning landmark studies—from real-space imaging of pair-density-wave states to explorations of electronic fluid flow[23-25].

## Results

As illustrated in Fig. 1a, the triple-**Q** CDW modulation distinctly overlays the topmost hexagonal lattice of Se atoms on the $NbSe_2$ surface. The local commensurate domain reveals a CDW unit cell matching three times the lattice periodicity in real space (inset of Fig. 1a). Remarkably, $C_{3v}$ point group symmetry is preserved in the CDW state, characterized by a modulation consisting of three symmetry-equivalent wavevectors oriented at intervals of 120°. Figure 1b shows the Fast Fourier Transform (FFT) of the STM image, where Bragg peaks and 3×3 CDW peaks simultaneously emerge. We denote the positions and intensities of the CDW peaks by vectors $\pm\mathbf{d}_i$ ($i$ = 1,2,3) and scalars $P(\pm\mathbf{d}_i)$, respectively (see Supplementary Fig. 1). Since the CDW is a real physical observable, its Fourier components exhibit complex conjugate symmetry, which guarantees that $P(\mathbf{d}_i) = P(-\mathbf{d}_i)$ (Supplementary Note 1). We therefore restrict our discussion to the set $\{\mathbf{d}_i\}$. In 2H-$NbSe_2$, $\{\mathbf{d}_1, \mathbf{d}_2, \mathbf{d}_3\} = \left\{(0,1), \left(\frac{\sqrt{3}}{2}, \frac{1}{2}\right), \left(-\frac{\sqrt{3}}{2}, \frac{1}{2}\right)\right\} 4\sqrt{3}\pi/(9a)$, where $a$ is the lattice constant. The nearly equal intensities of all CDW peaks, i.e., $P(\mathbf{d}_1) \approx P(\mathbf{d}_2) \approx P(\mathbf{d}_3)$, indicate a symmetric triple-**Q** order. By incorporating a triple-**Q** CDW potential together with superconductivity into a tight-binding model simulation (Supplementary Note 1), we obtain the simulated real-space local density of states (LDOS) of $NbSe_2$ at $E = 1.3\Delta$, which exhibits a clear 3×3 charge order (Fig. 1c). The corresponding FFT (Fig. 1d) reproduces the $P(\mathbf{d}_1) = P(\mathbf{d}_2) = P(\mathbf{d}_3)$ pattern, demonstrating a triple-**Q** CDW order with $C_{3v}$ symmetry (Supplementary Fig. 5).

Applying a magnetic field parallel to a superconductor's surface induces a diamagnetic surface current, i.e., the Meissner effect. To verify this effect in our 2H-$NbSe_2$ sample, we track the evolution of the superconducting gap under an in-plane magnetic field. In Fig. 1e, the averaged superconducting gaps expand as the applied field strength increases, accompanied by the emergence of a finite filling at the Fermi energy. The positions of the coherence peaks shift to higher energies with field up to 0.4 T (Supplementary Fig. 6). From the field dependence of the superconducting gap, we confirm that the screening currents induced by the in-plane field generate a Doppler shift. Expanding to a wider energy range, the spectra exhibit no discernible field-induced changes in normal states (Fig. 1f), implying the applied field mainly affects Bogoliubov quasiparticles within the superconducting gap.

Next, we study the interplay between the CDW order and Bogoliubov quasiparticles. By performing spectroscopic imaging via energy-dependent d$I$/d$V$ maps, we obtain real-space LDOS distributions, and the corresponding FFTs further reveal the CDW vectors in momentum space as a function of energy. At zero magnetic field, the superconducting gap of $NbSe_2$ shows pronounced coherence peaks at ±1.28 mV (Fig. 2a). Below 0.5 mV, the conductance drops to zero, indicating a fully gapped state. A clear kink appears near 1 mV, possibly arising from two-band pairing or an anisotropic gap structure[26-29]. Figures 2b-g present the FFTs of d$I$/d$V$ maps at the indicated energies (see Supplementary Figs. 7 and 8 for the full dataset). Based on these reciprocal space patterns, one can find that the CDW modulations superimposed on the normal states (3 mV) and Bogoliubov quasiparticles (0.6-1.2 mV) both exhibit 3×3 orders with equal $P(\mathbf{d}_i)$, indicative of preserved $C_{3v}$ symmetry. Within the fully gapped region (≤ 0.5 mV) where the LDOS vanishes entirely, no CDW signal is detected.

Remarkably, Bogoliubov quasiparticle excitations at the precursor CDW vectors exhibit drastically distinct behavior under in-plane fields. Upon applying a 100 mT field along the Γ-M direction, the coherence peaks shift to ±1.4 mV, while the gap edge moves to 0.2 mV (Fig. 2h), consistent with a Doppler shift induced by the screening current. We subsequently carried out spectroscopic imaging at various energies. As shown in Figs. 2i-k, the FFTs of the d$I$/d$V$ maps measured at $E$ = 0.2, 0.5 and 0.6 meV exhibit two CDW peaks significantly enhanced along the field direction, while the other four are suppressed. We obtain the pattern $P(\mathbf{d}_1) > P(\mathbf{d}_2) = P(\mathbf{d}_3)$ (here the vector parallel to the field direction is denoted as $\mathbf{d}_1$), which illustrates the symmetry breaking of the CDW from $C_{3v}$ at zero field to $C_s$ under the applied magnetic field. This enhancement remains robust throughout the 0.2-0.6 mV energy window (see Supplementary Figs. 9 and 10 for the full dataset). Outside this window, the CDW peaks retain to an equal intensity pattern at $E$ > 0.6 mV (Figs. 2l to 2n). Surface-vortex contributions are unlikely to account for the field-dependent CDW peak rearrangements, since the measurements were conducted in regions well separated from visible vortices (Supplementary Fig. 11). Given that the in-plane field predominantly modulates sub-gap Bogoliubov quasiparticles, we conclude that the field-induced enhancement of selected peaks is a direct consequence of the interplay between CDW order and Bogoliubov quasiparticles under the impact of supercurrent. In addition, the Bragg-peak intensities exhibit a field-direction dependence similar to that of the CDW (Figs. 2 i-k), as expected from standard Bloch physics. The associated supercurrent-induced Doppler shift reshapes the constant-energy contours inside the superconducting gap, selectively modulates Bragg scattering between states at $\mathbf{k}$ and $\mathbf{k}+\mathbf{G}$, thereby providing fundamental momentum-space evidence of the Doppler shift acting on Bogoliubov quasiparticles.

To further investigate the response of the CDW to different field orientations, we conducted spectroscopic imaging under 400 mT vector fields aligned with six high-symmetry directions. Figure 3 presents real- and momentum-space images acquired at $E$ = 0.6 mV, which resides in the energy window displaying CDW anisotropy. When the field is applied along the Γ-M directions, the real-space images display distinctly enhanced CDW modulations along the field directions, and the corresponding FFTs reveal marked intensity increase of two CDW peaks in those directions. Once the field is rotated to the Γ-K, the remaining four CDW peaks nearest the field direction exhibit pronounced intensity enhancements. In contrast, the CDW peaks in the FFTs of topographic images retain $C_{3v}$ symmetry upon rotating the field (Supplementary Fig. 12), demonstrating that the field-direction dependence of the CDW order originates from Bogoliubov quasiparticles.

**Discussion**

The lowering of the overall spectral weight symmetry to $C_s$ in momentum space is generically expected in superconductors due to the Doppler shift. The central question is the microscopic origin of the field-directional selectivity of the CDW response. To address this, we construct a real-space tight-binding model of $NbSe_2$, incorporating Ising type spin-orbital coupling, 3×3 triple-$\mathbf{Q}$ CDW order and superconductivity with a single s-wave pairing potential (see Methods section for details). The supercurrent effect is depicted by considering Peierls substitution, $\mathbf{p} \rightarrow \mathbf{p} + e\mathbf{A}$, where $\mathbf{p}$ is the Fermi momentum of $NbSe_2$, and $\mathbf{A}$ is the magnetic vector potential and is assumed to be perpendicular to the magnetic field $\mathbf{B}$ according to the London gauge (Supplementary Fig. 13). Particularly, in superconductors, $e\mathbf{A}$ represents the momentum of Cooper pair (i.e., the supercurrent), which is able to modify the pristine Bogoliubov quasiparticle dispersion into $E_{\pm}(\mathbf{k}) = \sqrt{(\epsilon_{\mathbf{k}} - \mu)^2 + \Delta^2} \pm e\mathbf{v}_{\mathbf{k}} \cdot \mathbf{A}$ . This framework allows us to track how supercurrents reshape low-energy quasiparticle contours and induce symmetry breaking in a CDW intertwined superconductor.

Our calculated results in Figs. 4a and b demonstrate that when the field is applied along the Γ-M direction on the $NbSe_2$ surface, Doppler shift induced modification of $E$-$\mathbf{k}$ dispersion emerges exclusively along the perpendicular Γ-K direction, and vice versa. When the Doppler shift satisfies $0 < E_{\mathrm{D}} < \Delta$, only half of the Bogoliubov quasiparticle branches are present on the constant energy contours within the interval $E_{\mathrm{D}} < E < \Delta$. In this scenario, the field-modified contour exhibits mirror symmetry, while the $C_{3v}$ symmetry present at zero field is broken. Figures 4c and d display the constant energy cuts taken at $E$ = 0.5Δ with indicated field directions. Beyond the confinement of spectral weight to one half of the Brillouin zone, two mutually perpendicular mirror planes $M_x$ and $M_y$ emerge, indicative of the resultant $C_s$ point group symmetry. For $E \geq \Delta$, the high density of states on the constant-energy contour renders Doppler-shift effect negligible, so the CDW orders retain $C_{3v}$ symmetry and all six peaks exhibit equal intensity, as in the zero-field case (Supplementary Figs. 16, 17).

The anisotropic modification of the Bogoliubov quasiparticle dispersion gives rise to a symmetry transition in real-space electronic patterns (Figs. 4e, f). In contrast to the homogeneous LDOS map at zero field (Fig. 1c) showing CDW modulation with $C_{3v}$ symmetry, our simulated real-space LDOS map under an in-plane field along Γ-M in Fig. 4e clearly displays a strip-shaped CDW pattern, consistent with $C_s$ symmetry. Applying the field along Γ-K results in two overlapping strip-shaped modulations (Fig. 4f). The corresponding FFTs of the real-space maps (Figs. 4g, h) directly capture the supercurrent-induced modification of the CDW pattern superimposed on the Bogoliubov quasiparticles. Namely, from the FFT line cuts in Figs. 4g and 4h, $P(\mathbf{d}_1)$ exceeds that of $P(\mathbf{d}_2)$ when the field is applied along Γ-M, whereas the opposite ordering is observed for a field along Γ-K. Thus, a pattern with $P(\mathbf{d}_1) > P(\mathbf{d}_2) = P(\mathbf{d}_3)$ [$P(\mathbf{d}_1) < P(\mathbf{d}_2) = P(\mathbf{d}_3)$] is obtained when the magnetic field is parallel to Γ-M (Γ-K) direction. Our simulations in real and reciprocal spaces reproduce the experimental data, confirming that the manipulation of the CDW patterns is indeed driven by the supercurrent effect.

To elucidate the relationship between magnetic-field orientation and the selective enhancement of CDW peaks, we perform a momentum-space analysis. At zero field, the CDW peaks at $\mathbf{d}_1$, $\mathbf{d}_2$ and $\mathbf{d}_3$ are degenerate in intensity by virtue of the underlying $C_{3v}$ symmetry. Under an applied magnetic field enforcing mirror symmetries $M_x$ or $M_y$, the CDW peaks $\mathbf{d}_2$ and $\mathbf{d}_3$ are symmetry-locked (i.e., $P(\mathbf{d}_2) = P(\mathbf{d}_3)$), whereas $\mathbf{d}_1$ is unconstrained by these symmetries and departs from the original $C_{3v}$ symmetry, enabling an independent amplitude (Supplementary Note 2). The corresponding CDW peak strengths, $P(\mathbf{d}_1)$ and $P(\mathbf{d}_2)$, are extracted through theoretical analysis (see Methods section for details):

$$P(\mathbf{d}_i) \propto \sum_{\mathbf{k}} \delta\left(E - E(\mathbf{k})\right)\Delta_{\mathrm{CDW}}\left(\frac{1}{\Delta E(\mathbf{k}+\mathbf{d}_i)} + \frac{1}{\Delta E(\mathbf{k}-\mathbf{d}_i)}\right) \tag{1}$$

where $i$ = 1, 2, $\Delta_{\mathrm{cdw}}$ is the CDW potential strength. $\sum_{\mathbf{k}} \delta\left(E - E(\mathbf{k})\right)$ selects all $\mathbf{k}$ on the constant energy contours, e.g., the red (blue) line in Fig. 4c (Fig. 4d). It's worth noting that $P(\mathbf{d}_i)$ is governed by the energy difference $\Delta E(\mathbf{k} \pm \mathbf{d}_i) = E(\mathbf{k}) - E(\mathbf{k} \pm \mathbf{d}_i)$. Accounting for the Doppler-shift-induced spectral asymmetry of the constant-energy contours, we divide the Brillouin zone into K-dominated and M-dominated halves (Figs. 4c, d). Under a field along Γ-M, the Doppler shift increases $\Delta E(\mathbf{k} \pm \mathbf{d}_2)$, which suppresses $P(\mathbf{d}_2)$ while leaving $P(\mathbf{d}_1)$ essentially unchanged (Fig. 4g). Conversely, with the field along Γ-K, $\Delta E(\mathbf{k}\pm\mathbf{d}_1)$ is enhanced, so that $P(\mathbf{d}_2) > P(\mathbf{d}_1)$ (Fig. 4h). This contrast confirms that the CDW's modulation of Bogoliubov quasiparticles at $\mathbf{d}_1$ and $\mathbf{d}_2$ respectively dominate the K- and M-dominated halves, underpinning the anisotropic CDW response to magnetic-field orientation.

Finally, it should be recognized that the applied magnetic fields in our experiment exceed $H_{c1}$ of $NbSe_2$, and vortices are therefore expected to be present in the sample. Although our STM measurements are performed in regions away from visible vortex cores at the surface, vortices are nevertheless present in the bulk. Under such conditions, where vortices are present and may overlap, the field distribution can become substantially more uniform than that expected from a Meissner-state screening profile, and the corresponding current distribution near the surface may therefore differ from that in a simple Meissner-screening picture. The pronounced symmetry lowering observed in our data may therefore include additional effects beyond the most straightforward Doppler-shift scenario. One possible explanation is that vortices are repelled from the surface, for example due to a strong surface-barrier effect, which would modify surface current distributions. Another possibility is that the Bogoliubov quasiparticles are intrinsically more sensitive to the in-plane magnetic field. Nevertheless, the Doppler shift is evidenced by the field evolution of the superconducting gap. The Doppler-shift framework therefore provides a natural and consistent explanation for the field-orientation-dependent redistribution of CDW peak intensities observed in our experiment.

In summary, our experiments and theoretical simulations uncover how supercurrent-induced Doppler shifts govern the interplay between superconductivity and CDW order in 2H-$NbSe_2$. We further demonstrate the capability to manipulate Bogoliubov quasiparticle excitations at the precursor CDW vectors, opening the possibility to the in-plane magnetic field manipulation of the Cooper pair density waves in a broader class of CDW intertwined superconductors. These insights also provide a potential route toward controlling quantum phases through engineered supercurrents and Doppler effects in superconductors intertwined with other quantum orders.

## Methods

**STM measurement:** The STM/STS measurements were performed in a scanning tunneling microscopes operating under ultra-high vacuum and equipped with a dilution refrigerator and a vector magnetic field ($H_x$: 2 T, $H_y$: 2 T, $H_z$: 9 T). The 2H-$NbSe_2$ single crystals were cleaved *in situ* at room temperature and then transferred immediately into STM head. All the measurements were carried out at a base temperature of 40 mK using tungsten tips treated with heating and silver decoration *in situ*. $dI/dV$ signals were acquired by a lock-in amplifier with modulation of 100 μV at 991 Hz, unless otherwise noted.

**Sample growth:** 2H-$NbSe_2$ single crystals were grown by the method of chemical vapor transport. High-purity Nb (powder, 99.99%) and Se (grain, 99.99%) were mixed in a molar ratio of 1:2 and sealed in an evacuated quartz tube with iodine as the transport agent. The quartz tube was subsequently placed in a dual-temperature zone tube furnace, with the raw material in the high-temperature zone and the crystal growth in the low-temperature zone. The furnace was initially heated to 573 K for 200 minutes and then maintained at this temperature for an additional 300 minutes. Following this, the high-temperature zone was further heated to 1065 K for 320 minutes, while the low-temperature zone was heated to 1025 K for 200 minutes. These respective temperatures were then sustained for a prolonged period of 15,000 minutes. Upon cooling to room temperature, the samples were successfully obtained on the low-temperature side of the tube.

**Numerical simulation**: The numerical simulations were performed using a real-space model electronic Hamiltonian $H_{\mathrm{e}}$ is

$$H_{\mathrm{e}} = H_{\mathrm{tb}} + H_{\mathrm{CDW}} \tag{2}$$

which represent the tight-binding electronic structure and the CDW potential, respectively. Considering that STM primarily probes the surface of $NbSe_2$ and the diamagnetic current predominantly distributes on the material's surface, we simplify the system simulation by constructing a two-dimensional tight-binding model Hamiltonian. According to first-principles calculations, the electronic states near the Γ-point and K-point are derived mainly from the Nb $\mathrm{d}_{z^2}$ and the $\mathrm{d}_{x^2-y^2}/\mathrm{d}_{xy}$ orbitals. As the CDW phenomena under consideration are not strongly influenced by valley physics, it is reasonable to concentrate on the Γ point in our analysis. Therefore, we consider only the $\mathrm{d}_{z^2}$ orbital on a triangular lattice. Accordingly, the tight-binding electronic Hamiltonian takes

$$H_{\rm tb} = \sum_{\mathbf{R}_i,\sigma} (\epsilon_{\rm d} - \mu) c^\dagger_{\mathbf{R}_i,\sigma} c_{\mathbf{R}_i,\sigma} + \sum_{<\mathbf{R}_i\sigma,\mathbf{R}_j\sigma>} (t_1 + \frac{1}{2i}\sigma\lambda v_{\mathbf{R}_i,\mathbf{R}_j}) c^\dagger_{\mathbf{R}_i\sigma} c_{\mathbf{R}_j\sigma} + \sum_{\ll\mathbf{R}_i\sigma,\mathbf{R}_j\sigma\gg} t_2\, c^\dagger_{\mathbf{R}_i\sigma} c_{\mathbf{R}_j\sigma}\,, \quad (3)$$

where $\mathbf{R}_i = (x, y)$ is the site vector of the real space, $\epsilon_d$ denotes the on-site energy of the orbital, $\mu$ is the chemical potential, $t_1$ and $t_2$ are spin-independent hopping amplitudes between the nearest and next-nearest neighboring sites respectively, $\lambda$ parameterizes the strength of the Ising spin-orbit coupling, $\sigma = \pm 1$ stands for spin-up and spin-down respectively, and $v_{\mathbf{R}_i,\mathbf{R}_j} = \pm 1$ is a bond dependent factor. Compared to the first principles calculations[30], we estimate the parameters as: $t_1 \approx 0.14$ eV, $t_2 \approx 0.11$ eV, $\lambda \approx 0.03$ eV. The lattice structure, as well as the corresponding band structure and Fermi surface, are presented in Supplementary Fig. 2.

To reproduce the experimentally observed 3×3 triple-**Q** CDW real-space pattern, the CDW-induced potential Hamiltonian takes $H_{\rm CDW} = \sum_{\mathbf{R}_i\sigma} \Delta_{\rm CDW}(\mathbf{R}_i)\, c^\dagger_{\mathbf{R}_i\sigma} c_{\mathbf{R}_i\sigma}$ , where

$$\begin{aligned}\Delta_{\rm CDW}(\mathbf{R}_i) &= \sum_{\mathbf{d}_1,\mathbf{d}_2,\mathbf{d}_3} 2A_0 cos(\mathbf{d}_i \cdot \mathbf{r} + \phi_i) \\ &= 2A_0\cos\left(\frac{4\sqrt{3}\pi}{9a}y - \frac{2\pi}{3}\right) + 2A_0\cos\left(\frac{2\pi}{3a}x + \frac{2\sqrt{3}\pi}{9a}y + \frac{2\pi}{3}\right) + 2A_0\cos\left(-\frac{2\pi}{3a}x + \frac{2\sqrt{3}\pi}{9a}y\right), \\ &\phi_1 = -\frac{2\pi}{3}, \quad \phi_2 = \frac{2\pi}{3}, \quad \phi_3 = 0\,. \end{aligned} \quad (4)$$

When the system becomes superconducting, the corresponding Bogoliubov-de Gennes (BdG) Hamiltonian is

$$H_{\rm BdG} = \begin{pmatrix} H_{\rm e} & H_{\rm pairing} \\ H^\dagger_{\rm pairing} & -H^*_{\rm e} \end{pmatrix}, \quad H_{paring} = \Delta_{\rm sc} \sum_{\mathbf{R}_i} c^\dagger_{\mathbf{R}_i\uparrow} c^\dagger_{\mathbf{R}_i\downarrow}, \quad (5)$$

where $\Delta_{\rm sc}$ is the s-wave superconducting gap.

An applied magnetic field parallel to the *x-y* plane induces the diamagnetic current at the top and bottom surfaces due to the Meissner effect. For the slab geometry shown in Supplementary Fig. 13, the supercurrent **J** and the vector potential **A** are related by the London equation, $\mathbf{J} = -\frac{n_s e^2}{m^*}\mathbf{A}$, within the London gauge ($\nabla \cdot \mathbf{A} = 0$). Combined with Maxwell's equations, the vector potential in a superconductor satisfies $\nabla^2 \mathbf{A} = \frac{\mathbf{A}}{\lambda_{\rm L}^2}$, where $\lambda_{\rm L}$ is the London penetration length. When the magnetic field is applied along the Γ-M direction, based on the boundary conditions and choose the Landau gauge $\partial_z A_x\left(z = \frac{d}{2}\right) = B,\ \partial_z A_x(z = -\frac{d}{2}) = B$ , we can get the vector potential as $\mathbf{A} =$ $(A_x, 0{,}0) = (B\lambda_\mathrm{L}\frac{\sinh\left(\frac{z}{\lambda_\mathrm{L}}\right)}{\cosh\left(\frac{d}{2\lambda_\mathrm{L}}\right)}, 0{,}0)$. Based on the Peierls substitution, the presence of the vector potential modifies the hopping amplitude in $H_{tb}$ as

$$\tilde{t}_{ij} \rightarrow t_{ij} e^{-i\frac{e}{\hbar}\int_{\mathbf{R}_i}^{\mathbf{R}_j}\mathbf{A}\cdot d\boldsymbol{l}} = t_{ij} e^{-i\frac{e}{\hbar}B\lambda_\mathrm{L} a \tanh\frac{d}{2\lambda_\mathrm{L}}(x_j - x_i)} = t_{ij} e^{-i\phi_{ij}}, \tag{6}$$

where $\phi_{ij} = \frac{e}{\hbar}B\lambda_\mathrm{L} a \tanh\frac{d}{2\lambda_\mathrm{L}}(x_j - x_i)$ denotes the phase during the hopping process; therefore, in the tight-binding Hamiltonian $H_\mathrm{tb}$, the hopping terms are modified as

$$\sum_{<\mathbf{R}_i\sigma, \mathbf{R}_j\sigma>} (t_1 + \frac{1}{2i}\sigma\lambda v_{\mathbf{R}_i,\mathbf{R}_j}) c^\dagger_{\mathbf{R}_i\sigma} c_{\mathbf{R}_j\sigma} \rightarrow \sum_{<\mathbf{R}_i\sigma, \mathbf{R}_j\sigma>} (t_1 + \frac{1}{2i}\sigma\lambda v_{\mathbf{R}_i,\mathbf{R}_j}) e^{-i\phi_{ij}} c^\dagger_{\mathbf{R}_i\sigma} c_{\mathbf{R}_j\sigma},$$

$$\sum_{\ll\mathbf{R}_i\sigma, \mathbf{R}_j\sigma\gg} t_2\, c^\dagger_{\mathbf{R}_i\sigma} c_{\mathbf{R}_j\sigma} \rightarrow \sum_{\ll\mathbf{R}_i\sigma, \mathbf{R}_j\sigma\gg} t_2\, e^{-i\phi_{ij}} c^\dagger_{\mathbf{R}_i\sigma} c_{\mathbf{R}_j\sigma}, \tag{7}$$

The derivation follows the same procedure for a magnetic field applied along the Γ-K direction. To calculate the real space LDOS for the large system sizes required, we employ the Kernel Polynomial Method[31], an efficient linear-scaling algorithm. The real-space LDOS then takes

$$\begin{aligned} \rho_E(\mathbf{R}_i) &= \sum_n \delta\,(E - E_n)\langle\mathbf{R}_i|\psi_n\rangle\langle\psi_n|\mathbf{R}_i\rangle \\ &= \frac{1}{\pi\sqrt{1-\tilde{E}^2}}\left(g_0 + 2\sum_{n=1}^{N} g_n\,\langle\mathbf{R}_i|T_n(\widetilde{H})|\mathbf{R}_i\rangle T_n(\tilde{E})\right). \end{aligned} \tag{8}$$

Here, $E_n$ represents the eigenenergy of the system, and $\psi_n$ denotes the corresponding eigenstate. $T_n$ are the Chebyshev polynomials of the first kind. $\widetilde{H}$ and $\tilde{E}$ refer to the rescaled Hamiltonian and its corresponding eigen energies, respectively. The rescaling is given by $\widetilde{H} = \frac{H - \frac{E_\mathrm{max}+E_\mathrm{min}}{2}}{\frac{E_\mathrm{max}-E_\mathrm{min}}{2}}$ which ensures that the energy spectrum lies within the interval $[-1,1]$, a necessary condition for performing the Chebyshev expansion. The kernel polynomial $g_n$ is used to dampen the Gibbs oscillations of the partial sum. In the following numerical calculations, the system consists of a 100×100×3 lattice. The kernel we choose is the Jackson kernel and the number of Chebyshev polynomial moments is set to 2700. We use the random vector method to estimate the LDOS, with 2000 random vectors. The above numerical simulations are primarily carried out using the Kwant package[32] in the Python programming language.

**Symmetry analysis:** To understand the observed CDW peaks, we perform a symmetry analysis by treating the CDW potential as a perturbation. Note that, even in the presence of the magnetic field, the two-dimensional Hamiltonian without CDW potential still preserves the translation symmetry, and its eigenstates are simply Bloch states denoted as $|\mathbf{k}\rangle$. The CDW potential will break the translation symmetry and couples different $|\mathbf{k}\rangle$ with $|\mathbf{k}\pm\mathbf{d}_1\rangle$,$|\mathbf{k}\pm\mathbf{d}_2\rangle$, and $|\mathbf{k}\pm\mathbf{d}_3\rangle$ due to its specific form as shown in Eq. (4). According to first-order perturbation theory, the eigenstate with CDW becomes $|\psi(\mathbf{k})\rangle = a_{0,\mathbf{k}}|\mathbf{k}\rangle + \sum_{l=1}^{6} a_{l,\mathbf{k}}\,|\mathbf{k}+\mathbf{d}_l\rangle$ with $l\in\{1,2,3,4,5,6\}$ (The distribution of $\mathbf{d}_l$ can be seen in Supplementary Fig.1, where they satisfy the relations $\mathbf{d}_4 = -\mathbf{d}_1, \mathbf{d}_5 = -\mathbf{d}_2, \mathbf{d}_6 = -\mathbf{d}_3$) and $a_{0,\mathbf{k}} \gg |a_{l,\mathbf{k}}|$ the coefficients up to the first order perturbation (Supplementary note 1). Here, for the sake of convenience in the derivation, $l$ is taken from 1 to 6, which differs from the definition in the main text where $\mathbf{d}_i$ is indexed with $i$ = 1 to 3. The corresponding real-space LDOS can then be expressed as:

$$\begin{aligned}\rho_E(\mathbf{R}_i) &= \sum_{\mathbf{k}} \delta\,(E-E(\mathbf{k}))\langle\mathbf{R}_i|\psi(\mathbf{k})\rangle\langle\psi(\mathbf{k})|\mathbf{R}_i\rangle \\ &= \sum_{\mathbf{k}} \delta\left(E-E(\mathbf{k})\right)\langle\mathbf{R}_i|(a_{0,\mathbf{k}}|\mathbf{k}\rangle + \sum_{l=1}^{6} a_{l,\mathbf{k}}\,|\mathbf{k}+\mathbf{d}_l\rangle) \\ &\qquad (a_{0,\mathbf{k}}^*\langle\mathbf{k}| + \sum_{m=1}^{6} a_{m,\mathbf{k}}^*\,\langle\mathbf{k}+\mathbf{d}_m|)|\mathbf{R}_i\rangle \\ &= \sum_{\mathbf{k}} \delta\,(E-E(\mathbf{k}))\frac{1}{N}(a_{0,\mathbf{k}}e^{i\mathbf{k}\cdot\mathbf{R}_i} + \sum_{l=1}^{6} a_{l,\mathbf{k}}e^{i(\mathbf{k}+\mathbf{d}_l)\cdot\mathbf{R}_i}) \\ &\qquad (a_{0,\mathbf{k}}^*e^{-i\mathbf{k}\cdot\mathbf{R}_i} + \sum_{m=0}^{6} a_{m,\mathbf{k}}^*e^{-i(\mathbf{k}+\mathbf{d}_m)\cdot\mathbf{R}_i}) \\ &\approx \sum_{\mathbf{k}} \delta\,(E-E(\mathbf{k}))\frac{1}{N}\{a_{0,\mathbf{k}}^2 + a_{0,\mathbf{k}}[\left(a_{1,\mathbf{k}}+a_{4,\mathbf{k}}^*\right)e^{i\mathbf{d}_1\cdot\mathbf{R}_i} \\ &\qquad +\left(a_{2,\mathbf{k}}+a_{5,\mathbf{k}}^*\right)e^{i\mathbf{d}_2\cdot\mathbf{R}_i} + \left(a_{3,\mathbf{k}}+a_{6,\mathbf{k}}^*\right)e^{i\mathbf{d}_3\cdot\mathbf{R}_i} + c.c]\}.\end{aligned} \tag{9}$$

Here we can always take $a_{0,\mathbf{k}}$ to be real so that

$$a_{l,\mathbf{k}} = \frac{\langle\mathbf{k}+\mathbf{d}_l|H_{\mathrm{CDW}}|\mathbf{k}\rangle}{E(\mathbf{k})-E(\mathbf{k}+\mathbf{d}_l)} = \frac{\Delta_{\mathrm{CDW}}e^{i\phi_l}}{\Delta E(\mathbf{k}+\mathbf{d}_l)}, \tag{10}$$

where $\phi_4 = -\phi_1$, $\phi_5 = -\phi_2$, and $\phi_6 = -\phi_3$. Notably, there are six terms in Eq. (9) which depend explicitly on the spatial position and arise from the coupling among different momentum states induced by the CDW potential.

To analyze the momentum-space structure of these patterns, we compute the Fourier transform of the LDOS with given $\mathbf{Q}$ as

$$
\begin{aligned}
P_E(\mathbf{Q}) &= \sum_{\mathbf{R}_i} \rho_E(\mathbf{R}_i)\, e^{-i\mathbf{Q}\cdot\mathbf{R}_i} \\
&= \sum_{\mathbf{R}_i} e^{-i\mathbf{Q}\cdot\mathbf{R}_i} \sum_{\mathbf{k}} \delta\left(E - E(\mathbf{k})\right) \frac{1}{N} \{|a_{0,\mathbf{k}}|^2 + [(a_{0,\mathbf{k}} a_{4,\mathbf{k}}^* + a_{1,\mathbf{k}} a_{0,\mathbf{k}}^*) e^{i\mathbf{d}_1\cdot\mathbf{R}_i} \\
&\quad + (a_{0,\mathbf{k}} a_{5,\mathbf{k}}^* + a_{2,\mathbf{k}} a_{0,\mathbf{k}}^*) e^{i\mathbf{d}_2\cdot\mathbf{R}_i} + (a_{0,\mathbf{k}} a_{3,\mathbf{k}}^* + a_{6,\mathbf{k}} a_{0,\mathbf{k}}^*) e^{i\mathbf{d}_3\cdot\mathbf{R}_i} + c.c]\} \\
&= \sum_{\mathbf{k}} \delta\left(E - E(\mathbf{k})\right) \{a_{0,\mathbf{k}}^2 \delta_{0,\mathbf{Q}} + a_{0,\mathbf{k}} [(a_{1,\mathbf{k}} + a_{4,\mathbf{k}}^*) \delta_{\mathbf{d}_1,\mathbf{Q}} + (a_{2,\mathbf{k}} + a_{5,\mathbf{k}}^*) \delta_{\mathbf{d}_2,\mathbf{Q}} \\
&\quad + (a_{3,\mathbf{k}} + a_{6,\mathbf{k}}^*) \delta_{\mathbf{d}_3,\mathbf{Q}} + c.c]\}.
\end{aligned}
\tag{11}
$$

According to Eq. (10), the six position-dependent terms correspond to six peaks in momentum space located at $\mathbf{d}_l$ with the form

$$
|P_E(\mathbf{d}_l)| = \sum_{\mathbf{k}} \delta\left(E - E(\mathbf{k})\right) a_{0,\mathbf{k}} \Delta_{\mathrm{CDW}} \left( \frac{1}{\Delta E(\mathbf{k} + \mathbf{d}_l)} + \frac{1}{\Delta E(\mathbf{k} - \mathbf{d}_l)} \right) \tag{12}
$$

Without applying the magnetic field, the system has $C_3$ rotational symmetry (an element of $C_{3v}$ point group). Therefore, $|P_E(\mathbf{d}_l)|$ for $l \in \{1, 2, 3, 4, 5, 6\}$ possess equal values because $\mathbf{d}_l$ are transformed to each other under $C_3$ rotation (see details in Supplementary note 1). When applying the magnetic field, the Meissner current breaks the $C_3$ rotational symmetry so that the $|P_E(\mathbf{d}_l)|$ have not to take the same values (see details in Supplementary note 2).

**Data availability**

The main figure data generated in this study have been deposited in the Zenodo database at https://doi.org/10.5281/zenodo.20091865. All the raw data generated during this study and the datasets analyzed in this study are available from the corresponding authors upon request.

**Code availability**

The calculations were performed using the open-source packages Kwant[32]. The codes that support the findings of this study are available from the corresponding author upon request.

## Acknowledgments

We thank V. Madhavan and H. Ma for helpful discussion. H.Z. discloses support for the research of this work from the National Natural Science Foundation of China [grant number 92365302], the Ministry of Science and Technology of China [grant number 2025YFE0201200], the Science and Technology Commission of Shanghai Municipality [grant number 24LZ1401000], and Cultivation Project of Shanghai Research Center for Quantum Sciences [grant number LZPY2024-04]. J.J. discloses support for the research of this work from the National Natural Science Foundation of China [grant number 12488101], the Science and Technology Commission of Shanghai Municipality [grant numbers 2019SHZDZX01 and 20QA1405100], and innovation program for Quantum Science and Technology [grant number 2021ZD0302500]. Z.Z. discloses support for the research of this work from the National Natural Science Foundation of China [grant number 12104292]. S.W. discloses support for the research of this work from the National Natural Science Foundation of China [grant number 92577203]. Y.L. discloses support for the research of this work from the National Natural Science Foundation of China [grant number 12474156]. X.L. discloses support for the research of this work from the National Natural Science Foundation of China [grant numbers 22325203 and 52102336], the Science and Technology Commission of Shanghai Municipality [grant number 24LZ1400800], Cultivation Project of Shanghai Research Center for Quantum Sciences [grant number LZPY2024-05], and innovation program for Quantum Science and Technology [grant number 2021ZD0302700]. W.C., D.L., P.H., Y.Y., Q.Y., S.X., R.X., X.C., J.S., D.G., C.L., and Z.X. declare no relevant funding.

## Author contributions

Z.Z., H.Z. and J.J. conceived project; Z.Z. and D.L. conducted the STM measurement with the help of P. H., Q.Y., S.X., R.X., X.C., J.S., D.G., Y.L., S.W. and C.L.; W.C. and X.L. performed the numerical simulation and symmetry analysis; Y.Y., and Z.X. prepare the $NbSe_2$ samples; Z.Z., W.C., X.L., H.Z. and J.J. wrote the manuscript, with input from all authors. All authors discussed the results, interpretation, and conclusions.

## Competing interests

The authors declare no competing interests.

**Figures**

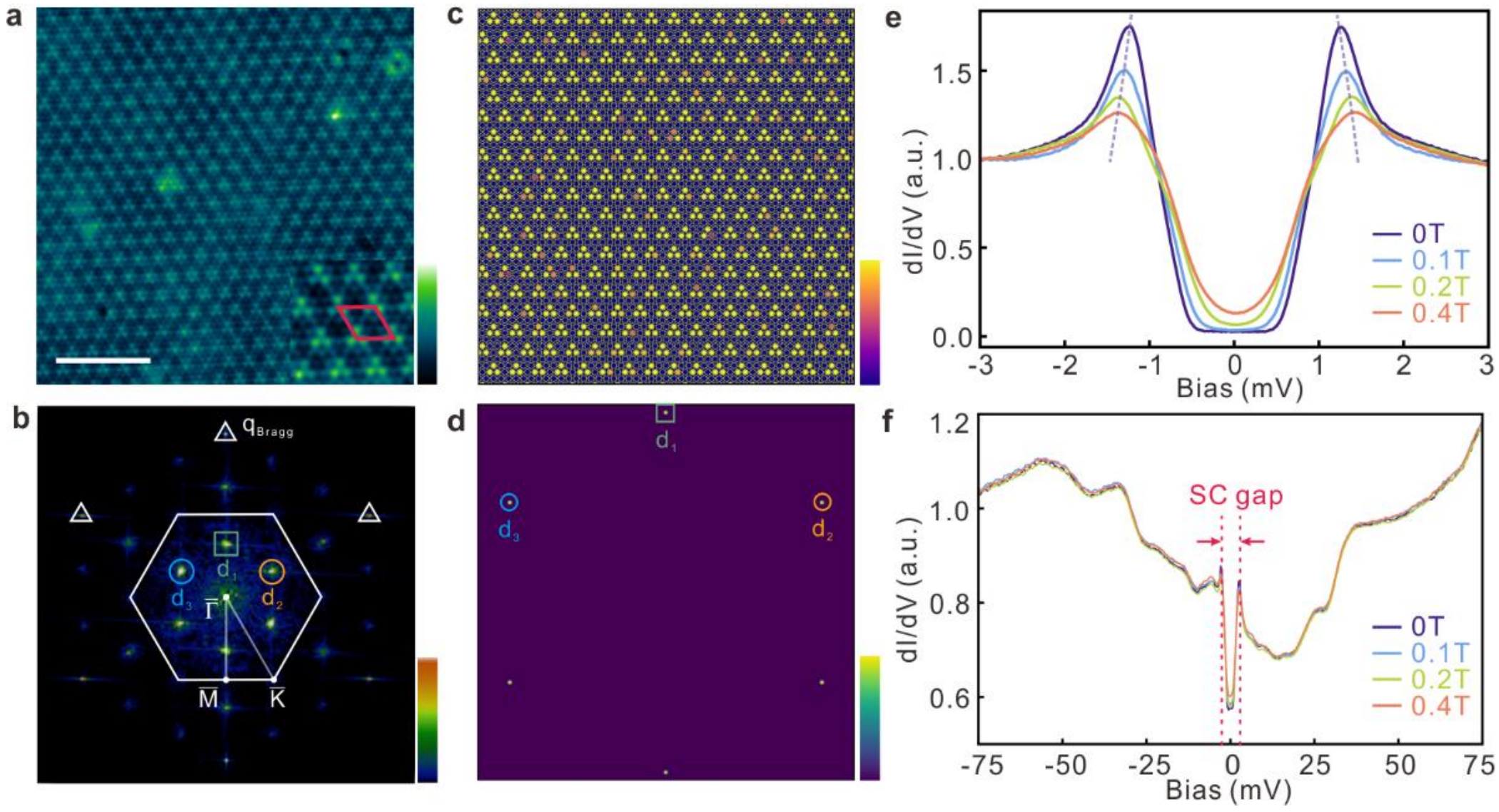

**Figure 1. CDW order and Doppler-shifted superconducting spectra in 2H-$NbSe_2$.** **a**, Atomically resolved topography showing 3×3 CDW superlattice superimposed on 1×1 atomic lattice ($V$ = 3 mV, $I$ = 100 pA). The inset shows a zoomed-in view, where the red rhombus marks the CDW unit cell. Scale bar, 5 nm. **b**, Fast Fourier transformed (FFT) of the STM image shown in **a**. The Bragg peaks are indicated by white triangles. The blue and orange circles and the green square mark the CDW peaks. The first Brillouin zone is outlined by a white hexagon. **c**, Simulated real-space local density of states (LDOS) with CDW pattern at $E$ = 1.3Δ. **d**, FFT of the simulated CDW pattern shown in **c**. The six CDW peaks exhibit equal intensities, indicating uniform modulation. **e**, Superconducting gap spectra of $NbSe_2$ measured under in-plane magnetic fields. With increasing field magnitude, the coherence peaks gradually shift outwards along with quasiparticle filling at the Fermi energy, indicative of a Doppler shift ($V$ = 3 mV, $I$ = 500 pA, $V_{mod}$ = 50 μV). **f**, Spectra measured in a wide energy range remain essentially unchanged regardless the field, confirming neglectable alteration in normal state electronic structure ($V$ = 100 mV, $I$ = 800 pA, $V_{mod}$ = 2 mV).

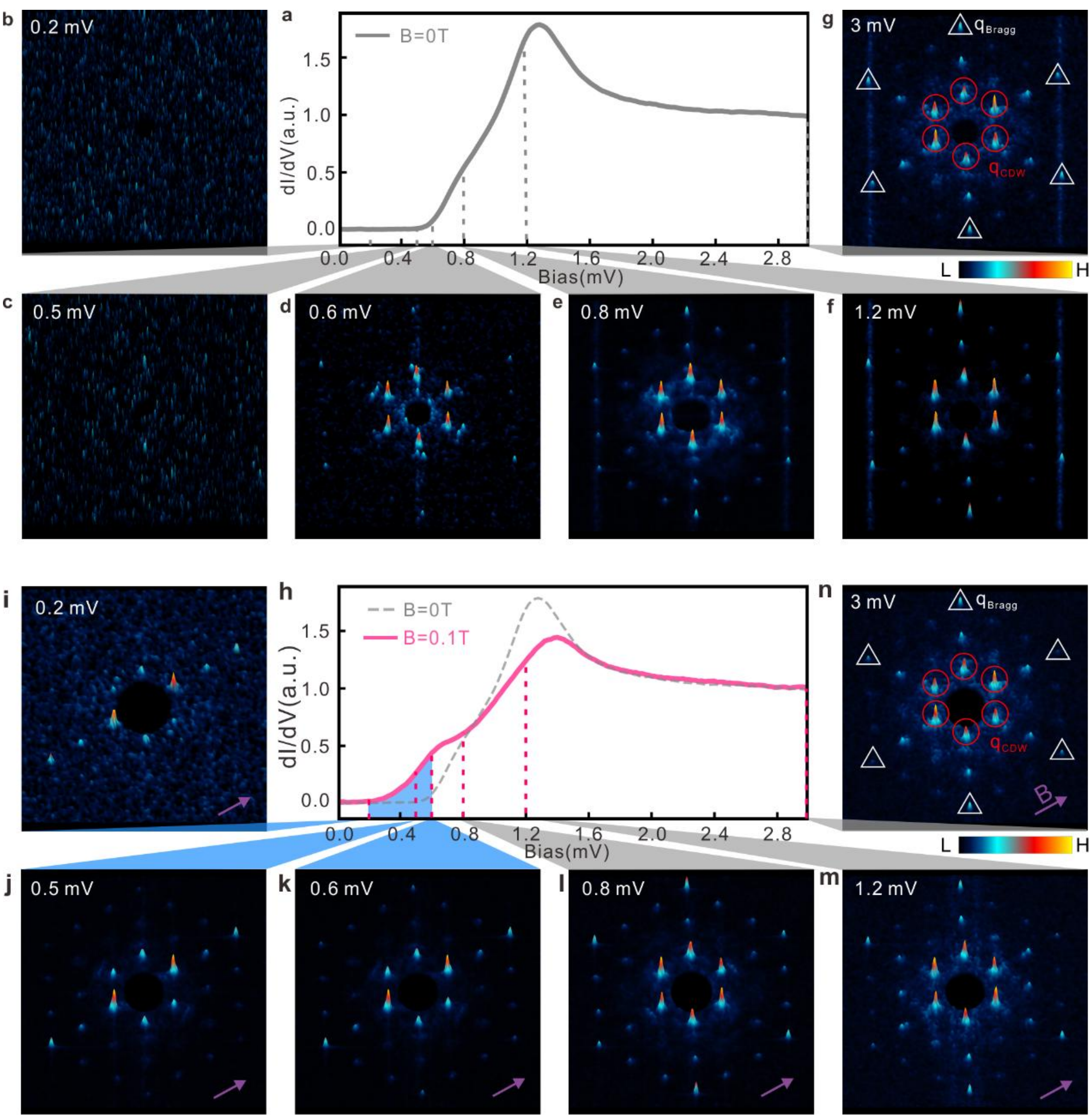

**Figure 2. Supercurrent effect of CDW modulations on Bogoliubov quasiparticles.** **a**, d$I$/d$V$ spectra taken under zero magnetic field ($V$ = 3 mV, $I$ = 500 pA). **b-g**, 3D representation of FFTed conductance maps ($V$ = 3 mV, $I$ = 500 pA) taken at indicated energies under zero field. Six CDW peaks (red circles) emerge above 0.5 mV with equal intensity. Bragg peaks are marked by white triangles. **h**, d$I$/d$V$ spectra (pink) measured under a 100 mT in-plane magnetic field along Γ-M direction ($V$ = 3 mV, $I$ = 500 pA). The dashed grey curve is the zero-field spectrum shown in **a**, overlaid for comparison. **i-n**, 3D representation of FFTed conductance maps ($V$ = 3 mV, $I$ = 500 pA) taken at indicated energies under a magnetic field of 100 mT. CDW and Bragg peaks are marked by red circles and white triangles, respectively. Two CDW peaks oriented parallel to the field direction are enhanced at 0.2mV, 0.5 mV and 0.6 mV, consistently observed across the blue-shaded energy range (see Supplementary Fig. 10). The CDW peaks exhibit comparable intensities at higher energies (0.8 mV, 1.2 mV and 3 mV) regardless of the magnetic field. Purple arrows denote the magnetic field directions.

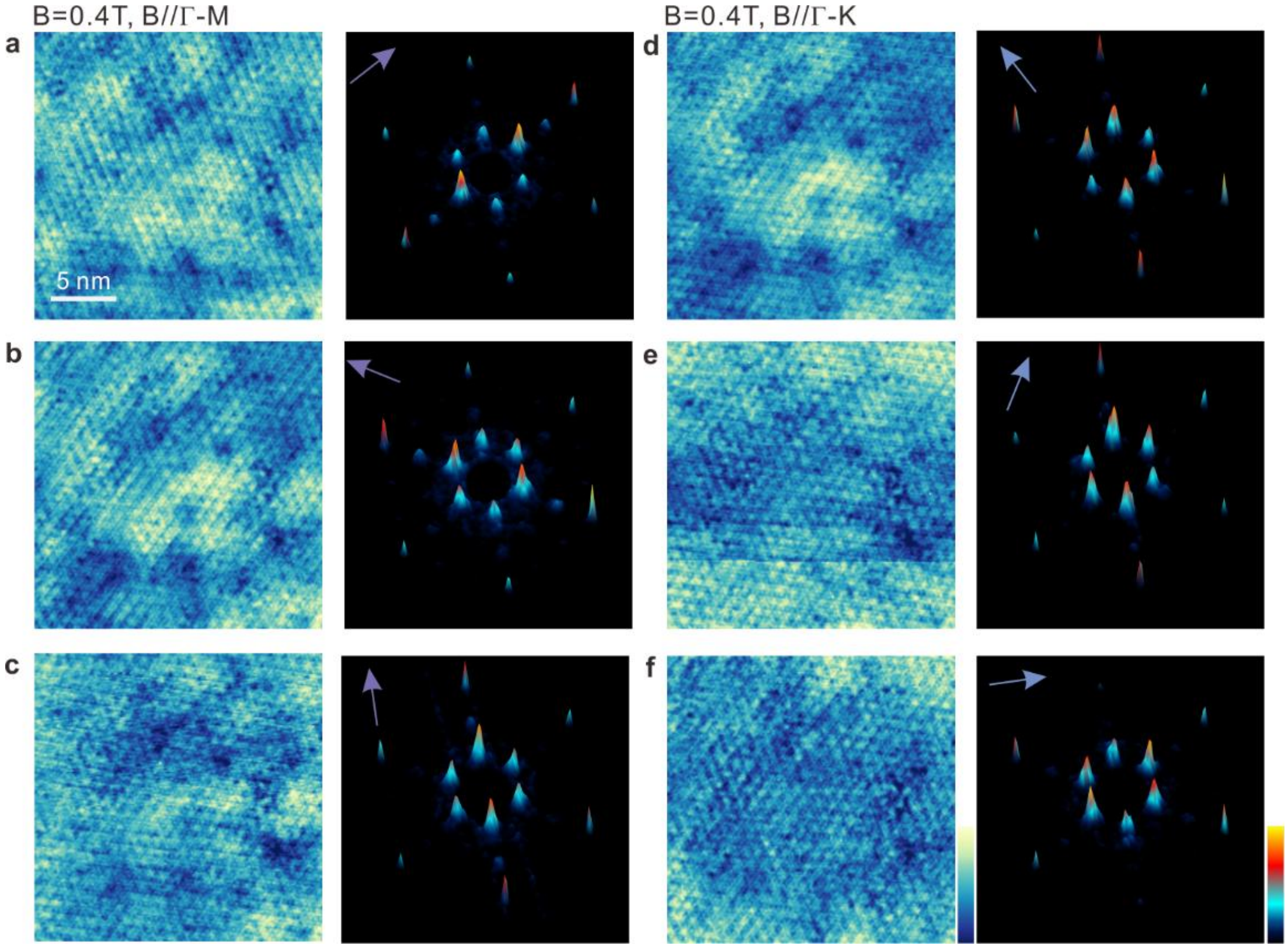

**Figure 3. Manipulation of the CDW modulations on Bogoliubov quasiparticles by in-plane magnetic fields. a-c**, Real-space d$I$/d$V$ maps and the corresponding 3D representation of FFT depicting the CDW modulations on Bogoliubov quasiparticles with magnetic field oriented along the Γ-M directions. Two CDW peaks aligned parallel to the field direction are selectively enhanced. **d-f**, Replicating **a-c** with the magnetic field applied along the Γ-K directions, where four CDW peaks closest to the field direction display markedly enhanced intensity. All data were acquired within the same 22×22 nm$^2$ area, at a bias voltage of 0.6 mV and under an in-plane magnetic field of 400 mT ($V$ = 2 mV, $I$ = 1 nA, $V_{\mathrm{mod}}$ = 50 μV). Field orientations are indicated by arrows.

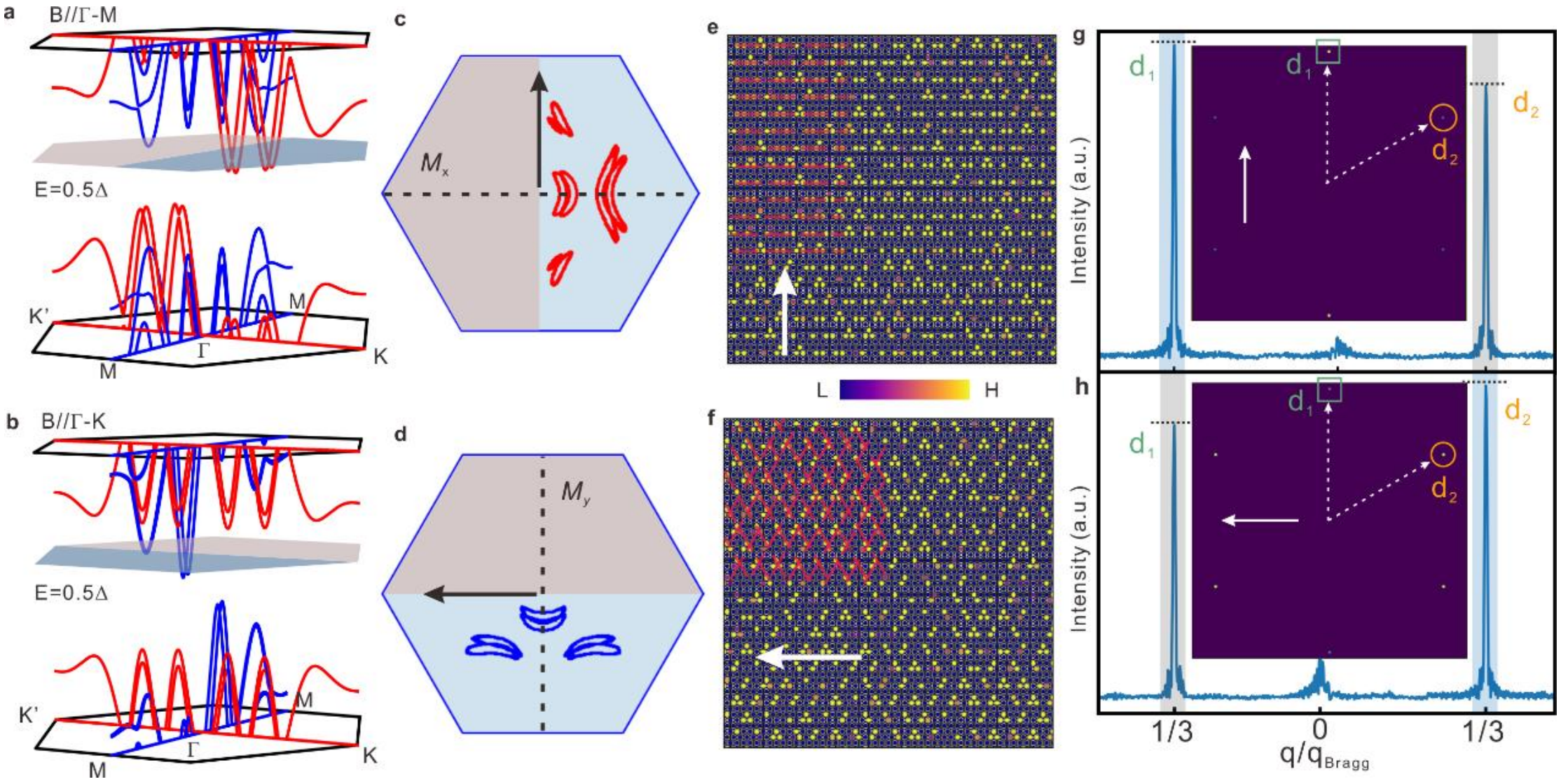

**Figure 4. Numerical simulations of supercurrent effect on CDW mediated patterns** **a, b,** Calculated Bogoliubov quasiparticle dispersions with fields applied along Γ-M and Γ-K directions, respectively. The Doppler shift develops perpendicular to the magnetic field and parallel to the direction of screening supercurrent. The blue and gray planes together represent constant-energy cuts at $E = 0.5\Delta$, where the blue plane highlights regions of the constant-energy contour that become populated by Doppler-shifted quasiparticles, while the gray plane remains gapped. The black hexagon denotes the Brillouin zone, with high-symmetry points indicated. **c, d,** Constant-energy contours at $E = 0.5\Delta$ extracted from **a** and **b**. Under field-induced supercurrent, Bogoliubov quasiparticles persist only on one half of the contour, lowering the system's symmetry down to $C_s$ with a single mirror plane (e.g., $M_x$ or $M_y$). **e, f,** Simulated real-space CDW-modulated LDOS patterns based on the constant-energy contours in **c** and **d.** The dominant CDW modulations are highlighted by red dashed lines. **g, h,** Line cuts taken along the dashed-arrow directions in the corresponding FFTs, shown as insets and derived from **e** and **f**. CDW peaks oriented parallel to the field direction exhibit pronounced enhancement, confirming the anisotropic CDW response to the supercurrent. All solid arrows indicate the in-plane field directions.

# Supplementary information for

## Supercurrent effect in a charge density wave intertwined superconductor

Zhen Zhu[1]†, Wei Cheng[2]†, Dang Liu[1]†, Pengyu Hu[1], Yi Yang[1], Qiaoyan Yu[1], Shasha Xue[1], Ruijun Xi[1], Xingsen Chen[1], Jice Sun[1], Dandan Guan[1,4,5], Yaoyi Li[1,4,5], Shiyong Wang[1,4,5], Canhua Liu[1,4,5], Zhuan Xu[3], Xin Liu[1,2,4]*, Hao Zheng[1,4,5]* & Jinfeng Jia[1,4,5,6,7]*

1. State Key Laboratory of Micro-nano Engineering Science, Tsung-Dao Lee Institute & School of Physics and Astronomy, Key Laboratory of Artificial Structures and Quantum Control (Ministry of Education), Shanghai Jiao Tong University, Shanghai 200240, China
2. School of Physics and Wuhan National High Magnetic Field Center, Huazhong University of Science and Technology, Wuhan, China
3. Department of Physics, Zhejiang University, Hangzhou, Zhejiang, China
4. Hefei National Laboratory, Hefei 230088, China
5. Shanghai Research Center for Quantum Sciences, 99 Xiupu Road, Shanghai 201315, China
6. Department of Physics, Southern University of Science and Technology, Shenzhen, China
7. Quantum Science Center of Guangdong-Hong Kong-Macao Greater Bay Area (Guangdong), Shenzhen, China

†These authors contributed equally to this work.

*Corresponding authors: haozheng1@sjtu.edu.cn ;phyliuxin@sjtu.edu.cn ; jfjia@sjtu.edu.cn

**Table of Contents**

**Supplementary Notes:**

**Supplementary Figures:**

**Supplementary Notes:**

When a 3×3 CDW potential is introduced, the Brillouin zone becomes smaller than the original one. The original and folded Brillouin zones are shown in Supplementary Fig. 1. The blue hexagon represents the original Brillouin zone, which applies to both the cases without and with a magnetic field, while the yellow hexagon denotes the folded Brillouin zone. As a result, states at different **k**-points within the original Brillouin zone get coupled, leading to the emergence of peak structures in the real-space LDOS.

To understand the observed CDW peaks, we perform a symmetry analysis by treating the CDW potential as a perturbation. we can get the relationship between the CDW peak intensities and the eigenstates on the Fermi surface (Eq. 11, Methods). As a result, we show that in the absence of a magnetic field, the system preserves threefold rotational symmetry, which ensures that the six peaks are equal in magnitude. However, when a magnetic field is applied, this rotational symmetry is broken, and only mirror symmetry remains. Consequently, the peaks along the direction of the magnetic field differ in intensity from those perpendicular to it.

To provide a more detailed explanation, the following analysis is divided into two parts: (i) CDW's modulation on LDOS of Bogoliubov quasiparticles in $NbSe_2$, (ii) Supercurrent effect on the CDW's modulation on LDOS in superconducting $NbSe_2$.

**Note 1: CDW's modulation on LDOS of Bogoliubov quasiparticles in $NbSe_2$**

We first consider the case where the CDW is included in the electronic part of the system. The Hamiltonian in this case is given by:

$$H_{\mathrm{e}} = H_{\mathrm{tb}} + H_{\mathrm{CDW}} \quad (S1)$$

By imposing periodic boundary conditions, we obtain:

$$\begin{aligned} H_{\mathrm{tb}} = \sum_{\mathbf{k}} [\, & 2t_1\cos(k_x) + 4t_1 \cos(\frac{\sqrt{3}}{2}k_y)\cos(\frac{1}{2}k_x) \\ & +2t_2\cos(\sqrt{3}k_y) + 4t_2\cos(\frac{3}{2}k_x)\cos(\frac{\sqrt{3}}{2}k_y) \\ & +\lambda\sigma(\sin(k_x) - 2\sin(\frac{1}{2}k_x)\cos(\frac{\sqrt{3}}{2}k_y))]c^{\dagger}_{\mathbf{k}\sigma}c_{\mathbf{k}\sigma} \end{aligned} \quad (S2)$$

where

$$H_{\mathrm{CDW}} = \sum_{\mathbf{d}_i}\sum_{\mathbf{k}\sigma} \Delta_{\mathrm{CDW}}\, e^{i\phi_i} c^{\dagger}_{\mathbf{k}\sigma}c_{(\mathbf{k}-\mathbf{d}_i)\sigma} + \Delta_{\mathrm{CDW}} e^{-i\phi_i} c^{\dagger}_{\mathbf{k}\sigma}c_{(\mathbf{k}+\mathbf{d}_i)\sigma}, \quad (S3)$$

$t_1$ and $t_2$ are spin-independent hopping amplitudes between the nearest and next-nearest neighboring sites respectively, $\lambda$ parameterizes the strength of the Ising spin-orbit coupling, $\sigma = \pm 1$ stands for spin-up and spin-down respectively. The band structure of $H_{\text{tb}}$ is shown in Supplementary Fig. 2. The experimentally observed 3×3 triple-**Q** CDW is characterized by the wave vectors $\mathbf{d}_i$ and phases $\phi_i$, which are defined as follows:

$$\mathbf{d}_1 = (0, \tfrac{4\sqrt{3}\pi}{9a})\ \mathbf{d}_2 = (\tfrac{2\pi}{3a}, \tfrac{2\sqrt{3}\pi}{9a})\ \mathbf{d}_3 = (-\tfrac{2\pi}{3a}, \tfrac{2\sqrt{3}\pi}{9a})\ \ \phi_1 = -\tfrac{2\pi}{3}\ \phi_2 = \tfrac{2\pi}{3}\ \phi_3 = 0 \qquad (S4)$$

where $a$ denotes the lattice constant. When the CDW is taken into account, the corresponding $C_3$-symmetric Fermi surface is shown in Supplementary Fig. 3.

We then consider the case of $T < T_c$ without a magnetic field, where superconductivity and CDW coexist in the system. The Hamiltonian in this case is given by:

$$H_{\text{sc}} = H_{\text{tb}} - \sum_{\mathbf{k}\sigma} \mu\, c_{\mathbf{k}\sigma}^{\dagger} c_{\mathbf{k}\sigma} + \sum_{\mathbf{k}} \Delta_{\text{sc}}\,(c_{\mathbf{k}\uparrow}^{\dagger} c_{-\mathbf{k}\downarrow}^{\dagger} + h.c) + H_{\text{CDW}} \qquad (S5)$$

Without the CDW, the eigenstates are simply Bloch states $|\mathbf{k}\rangle$. When the CDW is treated as a perturbation, it couples $|\mathbf{k}\rangle$ with $|\mathbf{k} \pm \mathbf{d}_1\rangle$, $|\mathbf{k} \pm \mathbf{d}_2\rangle$, and $|\mathbf{k} \pm \mathbf{d}_3\rangle$. According to first-order perturbation theory, the perturbed state becomes:

$$\begin{aligned} |\psi(\mathbf{k})\rangle &= a_{0,\mathbf{k}}|\mathbf{k}\rangle + a_{1,\mathbf{k}}|\mathbf{k} + \mathbf{d}_1\rangle + a_{2,\mathbf{k}}|\mathbf{k} + \mathbf{d}_2\rangle + a_{3,\mathbf{k}}|\mathbf{k} + \mathbf{d}_3\rangle \\ &\quad + a_{4,\mathbf{k}}|\mathbf{k} - \mathbf{d}_1\rangle + a_{5,\mathbf{k}}|\mathbf{k} - \mathbf{d}_2\rangle + a_{6,\mathbf{k}}|\mathbf{k} - \mathbf{d}_3\rangle \\ &= a_{0,\mathbf{k}}|\mathbf{k}\rangle + \sum_{l=1}^{6} a_{l,\mathbf{k}}\,|\mathbf{k} + \mathbf{d}_l\rangle \end{aligned} \qquad (S6)$$

With $l \in \{1, 2, 3, 4, 5, 6\}$, the distribution of $\mathbf{d}_l$ can be seen in Supplementary Fig. 1. Here, for the sake of convenience in the derivation, $l$ is taken from 1 to 6, which differs from the definition in the main text where $\mathbf{d}_i$ is indexed with $i$=1 to 3. The coefficients are given by:

$$a_{0,\mathbf{k}} = \frac{1}{C}$$

$$a_{1,\mathbf{k}} = \frac{1}{C}\frac{\langle \mathbf{k} + \mathbf{d}_1|H_{\text{CDW}}|\mathbf{k}\rangle}{E(\mathbf{k}) - E(\mathbf{k} + \mathbf{d}_1)} = \frac{1}{C}\frac{\Delta_{\text{CDW}} e^{-i\frac{2\pi}{3}}}{E(\mathbf{k}) - E(\mathbf{k} + \mathbf{d}_1)}$$

$$a_{2,\mathbf{k}} = \frac{1}{C}\frac{\langle \mathbf{k} + \mathbf{d}_2|H_{\text{CDW}}|\mathbf{k}\rangle}{E(\mathbf{k}) - E(\mathbf{k} + \mathbf{d}_2)} = \frac{1}{C}\frac{\Delta_{\text{CDW}} e^{i\frac{2\pi}{3}}}{E(\mathbf{k}) - E(\mathbf{k} + \mathbf{d}_2)} \qquad (S7)$$

Other coefficients follow similarly and $C$ denotes the normalization factor. The expression for the real-space LDOS is:

$$\rho_E(\mathbf{R}_i) = \sum_{\mathbf{k}} \delta\,(E - E(\mathbf{k}))\langle \mathbf{R}_i|\psi(\mathbf{k})\rangle\langle \psi(\mathbf{k})|\mathbf{R}_i\rangle \qquad (S8)$$

The $E(\mathbf{k})$ denotes the eigenenergy of the system, and $\psi(\mathbf{k})$ represents the corresponding eigenstate. In this case, we obtain:

$$
\begin{aligned}
\rho_E(\mathbf{R}_i) &= \sum_{\mathbf{k}} \delta\,(E - E(\mathbf{k}))\langle \mathbf{R}_i|\psi(\mathbf{k})\rangle\langle\psi(\mathbf{k})|\mathbf{R}_i\rangle \\
&= \sum_{\mathbf{k}} \delta\,\big(E - E(\mathbf{k})\big)\langle \mathbf{R}_i|(a_{0,k}|\mathbf{k}\rangle + \sum_{l=1}^{6} a_{l,\mathbf{k}}\,|\mathbf{k}+\mathbf{d}_l\rangle) \\
&\qquad (a_{0,\mathbf{k}}^{*}\langle\mathbf{k}| + \sum_{m=1}^{6} a_{m,\mathbf{k}}^{*}\,\langle\mathbf{k}+\mathbf{d}_m|)|\mathbf{R}_i\rangle \\
&= \sum_{\mathbf{k}} \delta\,(E - E(\mathbf{k}))\frac{1}{N}(a_{0,\mathbf{k}}e^{i\mathbf{k}\cdot\mathbf{R}_i} + \sum_{l=1}^{6} a_{l,\mathbf{k}}e^{i(\mathbf{k}+\mathbf{d}_l)\cdot\mathbf{R}_i}) \\
&\qquad (a_{0,\mathbf{k}}^{*}e^{-i\mathbf{k}\cdot\mathbf{R}_i} + \sum_{m=1}^{6} a_{m,\mathbf{k}}^{*}e^{-i(\mathbf{k}+\mathbf{d}_m)\cdot\mathbf{R}_i}) \\
&= \sum_{\mathbf{k}} \delta\,(E - E(\mathbf{k}))\frac{1}{N}[a_{0,\mathbf{k}}^{\mathbf{2}} + a_{0,\mathbf{k}}(\sum_{m=1}^{6} a_{m,\mathbf{k}}^{*}e^{-i\mathbf{d}_m\cdot\mathbf{R}_i}) \\
&\quad + a_{0,\mathbf{k}}^{*}(\sum_{l=1}^{6} a_{l,\mathbf{k}}e^{i\mathbf{d}_l\cdot\mathbf{R}_i}) + (\sum_{l=1}^{6} a_{l,\mathbf{k}}e^{i(\mathbf{k}+\mathbf{d}_l)\cdot\mathbf{R}_i})(\sum_{m=1}^{6} a_{m,\mathbf{k}}^{*}e^{-i(\mathbf{k}+\mathbf{d}_m)\cdot\mathbf{R}_i})]
\end{aligned}
\qquad (S9)
$$

Without the CDW, the eigenstates are simply Bloch states $|\mathbf{k}\rangle$. In this case, the LDOS in real space is a constant. When the CDW is taken into account, it couples different **k**-points, resulting in periodic oscillations in the LDOS. Due to $a_{0,\mathbf{k}} \gg |a_{l,\mathbf{k}}|$, we can keep terms up to the linear order of $a_{l,\mathbf{k}}$. We find six terms that depend explicitly on the spatial position:

$$
\begin{aligned}
\rho_E(\mathbf{R}_i) \approx \sum_{\mathbf{k}} \delta\,(E - E(\mathbf{k}))\frac{1}{N}\{&|a_{0,\mathbf{k}}|^2 + [\big(a_{0,\mathbf{k}}a_{4,\mathbf{k}}^{*} + a_{1,\mathbf{k}}a_{0,\mathbf{k}}^{*}\big)e^{i\mathbf{d}_1\cdot\mathbf{R}_i} \\
&+\big(a_{0,\mathbf{k}}a_{5,\mathbf{k}}^{*} + a_{2,\mathbf{k}}a_{0,\mathbf{k}}^{*}\big)e^{i\mathbf{d}_2\cdot\mathbf{R}_i} + \big(a_{0,\mathbf{k}}a_{6,\mathbf{k}}^{*} + a_{3,\mathbf{k}}a_{0,\mathbf{k}}^{*}\big)e^{i\mathbf{d}_3\cdot\mathbf{R}_i} + c.c]\}
\end{aligned}
\qquad (S10)
$$

Through Fourier transformation, we obtain that the six spatially dependent terms correspond to peak intensities in momentum space given by (Eq. 11, Methods):

$$
\begin{aligned}
P_E(\mathbf{d}_1) &= \sum_{\mathbf{k}} \delta\,(E - E(\mathbf{k}))\sum_{l,m} a_{l,\mathbf{k}}\,a_{m,\mathbf{k}}^{*}\delta_{\mathbf{q}_l-\mathbf{q}_m,\mathbf{d}_1} \\
&\approx \sum_{\mathbf{k}} \delta\,(E - E(\mathbf{k}))a_{0,\mathbf{k}}(a_{4,\mathbf{k}}^{*} + a_{1,\mathbf{k}}) \\
&= \sum_{\mathbf{k}} \delta\,\big(E - E(\mathbf{k})\big)\Delta_{\mathrm{CDW}}e^{-i\frac{2\pi}{3}}\frac{1}{C}\left(\frac{1}{\Delta E(\mathbf{k}+\mathbf{d}_1)} + \frac{1}{\Delta E(\mathbf{k}-\mathbf{d}_1)}\right)
\end{aligned}
$$

$$
\begin{aligned}
P_E(-\mathbf{d}_1) &= \sum_{\mathbf{k}} \delta\,(E - E(\mathbf{k})) \sum_{l,m} a_{l,\mathbf{k}}\, a^*_{m,\mathbf{k}} \delta_{\mathbf{q}_l - \mathbf{q}_m, -\mathbf{d}_1} \\
&\approx \sum_{\mathbf{k}} \delta\,(E - E(\mathbf{k})) a_{0,\mathbf{k}} (a_{4,\mathbf{k}} + a^*_{1,\mathbf{k}}) \\
&= \sum_{\mathbf{k}} \delta\,(E - E(\mathbf{k})) \Delta_{\mathrm{CDW}} e^{i\frac{2\pi}{3}} \frac{1}{C} \left( \frac{1}{\Delta E(\mathbf{k} + \mathbf{d}_1)} + \frac{1}{\Delta E(\mathbf{k} - \mathbf{d}_1)} \right) \\
P_E(\mathbf{d}_2) &= \sum_{\mathbf{k}} \delta\,(E - E(\mathbf{k})) \sum_{l,m} a_{l,\mathbf{k}}\, a^*_{m,\mathbf{k}} \delta_{\mathbf{q}_l - \mathbf{q}_m, \mathbf{d}_2} \\
&\approx \sum_{\mathbf{k}} \delta\,(E - E(\mathbf{k})) a_{0,\mathbf{k}} (a^*_{5,\mathbf{k}} + a_{2,\mathbf{k}}) \\
&= \sum_{\mathbf{k}} \delta\,(E - E(\mathbf{k})) \Delta_{\mathrm{CDW}} e^{i\frac{2\pi}{3}} \frac{1}{C} \left( \frac{1}{\Delta E(\mathbf{k} + \mathbf{d}_2)} + \frac{1}{\Delta E(\mathbf{k} - \mathbf{d}_2)} \right) \\
P_E(-\mathbf{d}_2) &= \sum_{\mathbf{k}} \delta\,(E - E(\mathbf{k})) \sum_{l,m} a_{l,\mathbf{k}}\, a^*_{m,\mathbf{k}} \delta_{\mathbf{q}_l - \mathbf{q}_m, -\mathbf{d}_2} \\
&\approx \sum_{\mathbf{k}} \delta\,(E - E(\mathbf{k})) a_{0,\boldsymbol{k}} (a_{5,\boldsymbol{k}} + a^*_{2,\boldsymbol{k}}) \\
&= \sum_{\mathbf{k}} \delta\,(E - E(\mathbf{k})) \Delta_{\mathrm{CDW}} e^{-i\frac{2\pi}{3}} \frac{1}{C} \left( \frac{1}{\Delta E(\mathbf{k} + \mathbf{d}_2)} + \frac{1}{\Delta E(\mathbf{k} - \mathbf{d}_2)} \right) \\
P_E(\mathbf{d}_3) &= \sum_{\mathbf{k}} \delta\,(E - E(\mathbf{k})) \sum_{l,m} a_{l,\mathbf{k}}\, a^*_{m,\mathbf{k}} \delta_{\mathbf{q}_l - \mathbf{q}_m, \mathbf{d}_3} \\
&\approx \sum_{\mathbf{k}} \delta\,(E - E(\mathbf{k})) a_{0,\mathbf{k}} (a^*_{6,\mathbf{k}} + a_{3,\mathbf{k}}) \\
&= \sum_{\mathbf{k}} \delta\,(E - E(\mathbf{k})) \Delta_{\mathrm{CDW}} \frac{1}{C} \left( \frac{1}{\Delta E(\mathbf{k} + \mathbf{d}_3)} + \frac{1}{\Delta E(\mathbf{k} - \mathbf{d}_3)} \right) \\
P_E(-\mathbf{d}_3) &= \sum_{\mathbf{k}} \delta\,(E - E(\mathbf{k})) \sum_{l,m} a_{l,\mathbf{k}}\, a^*_{m,\mathbf{k}} \delta_{\mathbf{q}_l - \mathbf{q}_m, -\mathbf{d}_3} \\
&\approx \sum_{\mathbf{k}} \delta\,(E - E(\mathbf{k})) a_{0,\mathbf{k}} (a_{6,\mathbf{k}} + a^*_{3,\mathbf{k}}) \qquad (S11) \\
&= \sum_{\mathbf{k}} \delta\,(E - E(\mathbf{k})) \Delta_{\mathrm{CDW}} \frac{1}{C} \left( \frac{1}{\Delta E(\mathbf{k} + \mathbf{d}_3)} + \frac{1}{\Delta E(\mathbf{k} - \mathbf{d}_3)} \right)
\end{aligned}
$$

We find that: $P_E(-\mathbf{d}_1) = P_E(\mathbf{d}_1)^*, P_E(-\mathbf{d}_2) = P_E(\mathbf{d}_2)^*, P_E(\mathbf{d}_3) = P_E(-\mathbf{d}_3)^*$, so we can know that $|P_E(-\mathbf{d}_1)| = |P_E(\mathbf{d}_1)|, |P_E(-\mathbf{d}_2)| = |P_E(\mathbf{d}_2)|, |P_E(-\mathbf{d}_3)| = |P_E(\mathbf{d}_3)|$, it means that the peak intensities of $\mathbf{d}_1$, $\mathbf{d}_2$, $\mathbf{d}_3$ is equal to$-\mathbf{d}_1$, $-\mathbf{d}_2$,$-\mathbf{d}_3$. This arises because the CDW potential is a real field with complex-conjugate symmetry.

Due to the system's C$_3$ rotational symmetry (an element of C$_{3v}$), we have $E(\mathbf{k}) = E(C_3\mathbf{k}) = E(C_3^2\mathbf{k})$. We can plot the energy contour at $E = \Delta$ as shown in Supplementary Fig. 4. The presence of C$_3$ rotational symmetry ensures that the six CDW peaks have equal intensity. In the following, we present a detailed proof of this result.

On the energy contour, the corresponding eigenstates are:

$$|\psi(C_3\mathbf{k})\rangle = b_{0,\mathbf{k}}|\mathbf{k}\rangle + b_{1,\mathbf{k}}|\mathbf{k}+\mathbf{d}_1\rangle + b_{2,\mathbf{k}}|\mathbf{k}+\mathbf{d}_2\rangle + b_{3,\mathbf{k}}|\mathbf{k}+\mathbf{d}_3\rangle$$
$$+b_{4,\mathbf{k}}|\mathbf{k}-\mathbf{d}_1\rangle + b_{5,\mathbf{k}}|\mathbf{k}-\mathbf{d}_2\rangle + b_{6,\mathbf{k}}|\mathbf{k}-\mathbf{d}_3\rangle$$

$$|\psi(C_3^2\mathbf{k})\rangle = c_{0,\mathbf{k}}|\mathbf{k}\rangle + c_{1,\mathbf{k}}|\mathbf{k}+\mathbf{d}_1\rangle + c_{2,\mathbf{k}}|\mathbf{k}+\mathbf{d_2}\rangle + c_{3,k}|\mathbf{k}+\mathbf{d_3}\rangle$$
$$+c_{4,\mathbf{k}}|\mathbf{k}-\boldsymbol{d}_1\rangle + c_{5,\mathbf{k}}|\mathbf{k}-\mathbf{d}_2\rangle + c_{6,\mathbf{k}}|\mathbf{k}-\mathbf{d}_3\rangle \qquad (S12)$$

In particular,

$$b_{1,\mathbf{k}} = \frac{1}{B}\frac{\Delta_{\mathrm{CDW}}e^{-\frac{2\pi}{3}i}}{E(C_3\mathbf{k}) - E(C_3\mathbf{k}+\mathbf{d}_1)} = \frac{1}{B}\frac{\Delta_{\mathrm{CDW}}e^{-\frac{2\pi}{3}i}}{E(\mathbf{k}) - E(\mathbf{k}-\mathbf{d}_3)} = \frac{C}{B}a_{6,\mathbf{k}}e^{-i\frac{2\pi}{3}} \qquad (S13)$$

where $B$ is another normalization constant. Similarly, we obtain:

$$b_{2,\mathbf{k}} = \frac{C}{B}a_{4,\mathbf{k}} \quad b_{3,\mathbf{k}} = \frac{C}{B}a_{2,\mathbf{k}}e^{i\frac{2\pi}{3}} \quad b_{4,\mathbf{k}} = \frac{C}{B}a_{3,\mathbf{k}}e^{i\frac{2\pi}{3}}$$
$$b_{5,\mathbf{k}} = \frac{C}{B}a_{1,\mathbf{k}} \quad b_{6,\mathbf{k}} = \frac{C}{B}a_{5,\mathbf{k}}e^{-i\frac{2\pi}{3}} \qquad (S14)$$

Normalization of the coefficients gives:

$$\sum_m b_{m,\mathbf{k}}^2 = 1 \rightarrow \frac{1}{B^2} + \frac{Ca_{6,\mathbf{k}}^2}{B^2} + \frac{Ca_{4,\mathbf{k}}^2}{B^2} + \frac{Ca_{2,\mathbf{k}}^2}{B^2} + \frac{Ca_{3,\mathbf{k}}^2}{B^2} + \frac{Ca_{1,\mathbf{k}}^2}{B^2} + \frac{Ca_{5,\mathbf{k}}^2}{B^2} = 1$$

$$\sum_m a_{m,\mathbf{k}}^2 = 1 \rightarrow \frac{1}{C^2} + a_{1,\mathbf{k}}^2 + a_{2,\mathbf{k}}^2 + a_{3,\mathbf{k}}^2 + a_{4,\mathbf{k}}^2 + a_{5,\mathbf{k}}^2 + a_{6,\mathbf{k}}^2 = 1$$

$$\rightarrow 1 + C^2a_{1,\mathbf{k}}^2 + C^2a_{2,\mathbf{k}}^2 + C^2a_{3,\mathbf{k}}^2 + C^2a_{4,\mathbf{k}}^2 + C^2a_{5,\mathbf{k}}^2 + C^2a_{6,\mathbf{k}}^2 = C^2 \qquad (S15)$$

Substituting this relation into the previous expression yields $C^2 = B^2$, indicating that the normalization constants are identical. From symmetry considerations, one can deduce the following relations among the coefficients:

$$b_{0,\mathbf{k}} = a_{0,\mathbf{k}} \quad b_{1,\mathbf{k}} = a_{6,\mathbf{k}}e^{-i\frac{2\pi}{3}} \quad b_{2,\mathbf{k}} = a_{4,\mathbf{k}} \quad b_{3,\mathbf{k}} = a_{2,\mathbf{k}}e^{i\frac{2\pi}{3}}$$
$$b_{4,\mathbf{k}} = a_{3,\mathbf{k}}e^{i\frac{2\pi}{3}} \quad b_{5,\mathbf{k}} = a_{1,\mathbf{k}} \quad b_{6,\mathbf{k}} = a_{5,\mathbf{k}}e^{-i\frac{2\pi}{3}}$$
$$c_{0,\mathbf{k}} = a_{0,\mathbf{k}} \quad c_{1,\mathbf{k}} = a_{5,\mathbf{k}} \quad c_{2,\mathbf{k}} = a_{3,\mathbf{k}}e^{i\frac{2\pi}{3}} \quad c_{3,\mathbf{k}} = a_{4,\mathbf{k}}e^{i\frac{2\pi}{3}}$$
$$c_{4,\mathbf{k}} = a_{2,\mathbf{k}} \quad c_{5,\mathbf{k}} = a_{6,\mathbf{k}}e^{-i\frac{2\pi}{3}} \quad c_{6,\mathbf{k}} = a_{1,\mathbf{k}}e^{-i\frac{2\pi}{3}} \qquad (S16)$$

Based on these relations, the contributions to the CDW from the Fermi surface states connected by $C_3$ symmetry can all be expressed using the same coefficient.

$$\begin{aligned}
P_E(\mathbf{d}_1) &= \sum_{\mathbf{k}} \delta\,(E - E(\mathbf{k}))[a_{0,\mathbf{k}}a_{4,\mathbf{k}}^* + a_{1,\mathbf{k}}a_{0,\mathbf{k}}^* + b_{0,\mathbf{k}}b_{4,\mathbf{k}}^* + b_{1,\mathbf{k}}b_{0,\mathbf{k}}^* + c_{0,\mathbf{k}}c_{4,\mathbf{k}}^* + c_{1,\mathbf{k}}c_{0,\mathbf{k}}^*] \\
&= \sum_{\mathbf{k}} \delta\left(E - E(\mathbf{k})\right)a_{0,\mathbf{k}}(a_{4,\mathbf{k}}^* + b_{4,\mathbf{k}}^* + c_{4,\mathbf{k}}^* + a_{1,\mathbf{k}} + b_{1,\mathbf{k}} + c_{1,\mathbf{k}}) \\
&= \sum_{\mathbf{k}} \delta\,(E - E(\mathbf{k}))a_{0,\mathbf{k}}(a_{4,\mathbf{k}}^* + a_{3,\mathbf{k}}^*e^{i\frac{2\pi}{3}} + a_{2,\mathbf{k}}^* + a_{1,\mathbf{k}} + a_{6,\mathbf{k}}e^{-i\frac{2\pi}{3}} + a_{5,\mathbf{k}}) \\
P_E(\mathbf{d}_2) &= \sum_{\mathbf{k}} \delta\,(E - E(\mathbf{k}))[a_{0,\mathbf{k}}a_{5,\mathbf{k}}^* + a_{0,\mathbf{k}}^*a_{2,\mathbf{k}} + b_{0,\mathbf{k}}b_{5,\mathbf{k}}^* + b_{0,\mathbf{k}}^*b_{2,\mathbf{k}} + c_{0,\mathbf{k}}c_{5,\mathbf{k}}^* + c_{0,\mathbf{k}}^*c_{2,\mathbf{k}}] \\
&= \sum_{\mathbf{k}} \delta\left(E - E(\mathbf{k})\right)a_{0,\mathbf{k}}(a_{5,\mathbf{k}}^* + b_{5,\mathbf{k}}^* + c_{5,\mathbf{k}}^* + a_{2,\mathbf{k}} + b_{2,\mathbf{k}} + c_{2,\mathbf{k}}) \\
&= \sum_{\mathbf{k}} \delta\left(E - E(\mathbf{k})\right)a_{0,\mathbf{k}}(a_{5,\mathbf{k}}^* + a_{1,\mathbf{k}}^* + a_{6,\mathbf{k}}^*e^{i\frac{2\pi}{3}} + a_{2,\mathbf{k}} + a_{4,\mathbf{k}} + a_{3,\mathbf{k}}e^{-i\frac{2\pi}{3}}) \\
&= P(\boldsymbol{d}_1)^* \\
P_E(\mathbf{d}_3) &= \sum_{\mathbf{k}} \delta\left(E - E(\mathbf{k})\right)[a_{0,\mathbf{k}}a_{6,\mathbf{k}}^* + a_{0,\mathbf{k}}^*a_{3,\mathbf{k}} + b_{0,\mathbf{k}}b_{6,\mathbf{k}}^* + b_{0,\mathbf{k}}^*b_{3,\mathbf{k}} + c_{0,\mathbf{k}}c_{6,\mathbf{k}}^* + c_{0,\mathbf{k}}^*c_{3,\mathbf{k}}] \\
&= \sum_{\mathbf{k}} \delta\,(E - E(\mathbf{k}))[a_{0,\mathbf{k}}(a_{6,\mathbf{k}}^* + b_{6,\mathbf{k}}^* + c_{6,\mathbf{k}}^* + a_{3,\mathbf{k}} + b_{3,\mathbf{k}} + c_{3,\mathbf{k}})] \\
&= \sum_{\mathbf{k}} \delta\left(E - E(\mathbf{k})\right)a_{0,\mathbf{k}}(a_{6,\mathbf{k}}^* + a_{5,\mathbf{k}}^*e^{-i\frac{2\pi}{3}} + a_{1,\mathbf{k}}^*e^{-i\frac{2\pi}{3}} + a_{3,\mathbf{k}} + a_{2,\mathbf{k}}e^{-i\frac{2\pi}{3}} + a_{4,\mathbf{k}}e^{-i\frac{2\pi}{3}}) \\
&= e^{-i\frac{2\pi}{3}}P_E(\mathbf{d}_1)^*
\end{aligned} \qquad (S17)$$

By comparing the coefficients above, we obtain the following relations:

$$P_E(\mathbf{d}_1) = P_E(\mathbf{d}_2)^* = e^{-i\frac{2\pi}{3}}P_E(\mathbf{d}_3)^* \qquad (S18)$$

This implies:

$$|P_E(\mathbf{d}_1)| = |P_E(\mathbf{d}_2)| = |P_E(\mathbf{d}_3)| \qquad (S19).$$

Using the numerical simulation methods detailed in the Methods section of the main text, we calculated the results after incorporating both superconductivity and CDW, as shown in Supplementary Fig. 5. From the band structures and the density of states (DOS) in Supplementary Figs. 5a-c, it is evident that a superconducting gap is opened in the system. Furthermore, we calculated the LDOS distributions at and outside the coherence peaks (Supplementary Figs. 5d, e), where 3×3 CDW modulations are clearly visible. By taking the FFT of the LDOS, we find that the six CDW-related peaks exhibit equal intensities (Supplementary Figs. 5f, g), which is consistent with the results of our preceding symmetry analysis.

**Note 2: Supercurrent effect on the CDW's modulation on LDOS in superconducting $NbSe_2$**

When the magnetic field is applied along the Γ-M direction (Supplementary Fig. 13):

$$H_{\mathrm{sc}}^{B}=H_{\mathrm{tb}}^{B}-\sum_{\mathbf{k}\sigma}\mu\, c_{\mathbf{k}\sigma}^{\dagger}c_{\mathbf{k}\sigma}+\sum_{\mathbf{k}}\Delta_{\mathrm{sc}}\, c_{\mathbf{k}\uparrow}^{\dagger}c_{-\mathbf{k}\downarrow}^{\dagger}+h.c+H_{\mathrm{CDW}} \qquad (S20)$$

where

$$\begin{aligned}H_{\mathrm{tb}}^{B}=\sum_{\mathbf{k}}[\,&2t_1\cos(k_x-\phi)+4t_1\cos(\frac{\sqrt{3}}{2}k_y)\cos(\frac{1}{2}k_x-\phi)+2t_2\cos(\sqrt{3}k_y)\\&+4t_2\cos(\frac{3}{2}k_x-\phi)\cos(\frac{\sqrt{3}}{2}k_y)+\lambda\sigma(\sin(k_x-\phi)-2\sin(\frac{1}{2}k_x)\cos(\frac{\sqrt{3}}{2}k_y))]c_{\mathbf{k}\sigma}^{\dagger}c_{\mathbf{k}\sigma}\end{aligned} \qquad (S21)$$

with $\phi=\frac{e}{\hbar}B\lambda_{\mathrm{L}}a\tanh\frac{d}{2\lambda_{\mathrm{L}}}$ is hopping phase (Eq. 6, Methods). At this time only the mirror symmetry with respect to the $x$-axis is preserved. The constant energy contour at $E$=0.5Δ in Supplementary Fig. 14 shows that, at this point, the Fermi surface only retains mirror symmetry with respect to the $x$-axis.

The eigenstate related to $\psi(\mathbf{k})$ by mirror reflection about the $x$-axis is

$$\begin{aligned}|\psi(\sigma_x\mathbf{k})\rangle=&f_{0,\mathbf{k}}|\mathbf{k}\rangle+f_{1,\mathbf{k}}|\mathbf{k}+\mathbf{d}_1\rangle+f_{2,\mathbf{k}}|\mathbf{k}+\mathbf{d}_2\rangle+f_{3,\mathbf{k}}|\mathbf{k}+\mathbf{d}_3\rangle\\&+f_{4,\mathbf{k}}|\mathbf{k}-\mathbf{d}_1\rangle+f_{5,\mathbf{k}}|\mathbf{k}-\mathbf{d}_2\rangle+f_{6,\mathbf{k}}|\mathbf{k}-\mathbf{d}_3\rangle\end{aligned} \qquad (S22)$$

In this case, we obtain the relationship between the coefficients can be derived as:

$$\begin{aligned}&f_{0,\mathbf{k}}=a_{0,\mathbf{k}}\quad f_{1,\mathbf{k}}=a_{4,\mathbf{k}}e^{i\frac{2\pi}{3}}\quad f_{2,\mathbf{k}}=a_{6,\mathbf{k}}e^{i\frac{2\pi}{3}}\quad f_{3,\mathbf{k}}=a_{5,\mathbf{k}}e^{i\frac{2\pi}{3}}\\&f_{4,\mathbf{k}}=a_{1,\mathbf{k}}e^{-i\frac{2\pi}{3}}\quad f_{5,\mathbf{k}}=a_{3,\mathbf{k}}e^{-i\frac{2\pi}{3}}\quad f_{6,\mathbf{k}}=a_{2,\mathbf{k}}e^{-i\frac{2\pi}{3}}\end{aligned} \qquad (S23)$$

It then follows that:

$$\begin{aligned}P_E(\mathbf{d}_1)&=\sum_{\mathbf{k}}\delta\,(E-E(\mathbf{k}))(a_{0,\mathbf{k}}a_{4,\mathbf{k}}^{*}+a_{0,\mathbf{k}}^{*}a_{1,\mathbf{k}}+f_{0,\mathbf{k}}f_{4,\mathbf{k}}^{*}+f_{0,\mathbf{k}}^{*}f_{1,\mathbf{k}})\\&=\sum_{\mathbf{k}}\delta\,(E-E(\mathbf{k}))a_{0,\mathbf{k}}(a_{4,\mathbf{k}}^{*}+f_{4,\mathbf{k}}^{*}+a_{1,\mathbf{k}}+f_{1,\mathbf{k}})\\&=\sum_{\mathbf{k}}\delta\left(E-E(\mathbf{k})\right)a_{0,\mathbf{k}}(a_{4,\mathbf{k}}^{*}+a_{1,\mathbf{k}}^{*}e^{i\frac{2\pi}{3}}+a_{1,\mathbf{k}}+a_{4,\mathbf{k}}e^{i\frac{2\pi}{3}})\end{aligned}$$

$$
\begin{aligned}
P_E(\mathbf{d}_2) &= \sum_{\mathbf{k}} \delta\left(E - E(\mathbf{k})\right)(a_{0,\mathbf{k}}a_{5,\mathbf{k}}^* + a_{0,\mathbf{k}}^*a_{2,\mathbf{k}} + f_{0,\mathbf{k}}f_{5,\mathbf{k}}^* + f_{0,\mathbf{k}}^*f_{2,\mathbf{k}}) \\
&= \sum_{\mathbf{k}} \delta\left(E - E(\mathbf{k})\right)a_{0,\mathbf{k}}(a_{5,\mathbf{k}}^* + f_{5,\mathbf{k}}^* + a_{2,\mathbf{k}} + f_{2,\mathbf{k}}) \\
&= \sum_{\mathbf{k}} \delta\left(E - E(\mathbf{k})\right)a_{0,\mathbf{k}}(a_{5,\mathbf{k}}^* + a_{3,\mathbf{k}}^*e^{i\frac{2\pi}{3}} + a_{2,\mathbf{k}} + a_{6,\mathbf{k}}e^{i\frac{2\pi}{3}})
\end{aligned}
$$

$$
\begin{aligned}
P_E(\mathbf{d}_3) &= \sum_{\mathbf{k}} \delta\left(E - E(\mathbf{k})\right)(a_{0,\mathbf{k}}a_{6,\mathbf{k}}^* + a_{0,\mathbf{k}}^*a_{3,\mathbf{k}} + f_{0,\mathbf{k}}f_{6,\mathbf{k}}^* + f_{0,\mathbf{k}}^*f_{3,\mathbf{k}}) \\
&= \sum_{\mathbf{k}} \delta\left(E - E(\mathbf{k})\right)a_{0,\mathbf{k}}(a_{6,\mathbf{k}}^* + f_{6,\mathbf{k}}^* + a_{3,\mathbf{k}} + f_{3,\mathbf{k}}) \\
&= \sum_{\mathbf{k}} \delta\left(E - E(\mathbf{k})\right)a_{0,\mathbf{k}}(a_{6,\mathbf{k}}^* + a_{2,\mathbf{k}}^*e^{i\frac{2\pi}{3}} + a_{3,\mathbf{k}} + a_{5,\mathbf{k}}e^{i\frac{2\pi}{3}}) \\
&= e^{i\frac{2\pi}{3}}P_E(\mathbf{d}_2)^*
\end{aligned}
\qquad (S24)
$$

From this, we obtain $P_E(\mathbf{d}_3) = e^{i\frac{2\pi}{3}}P_E(\mathbf{d}_2)^*$. It follows that the peak magnitudes of $\mathbf{d}_2$ and $\mathbf{d}_3$ are equal. At the same time, it can be seen that $P_E(\mathbf{d}_1)$ is independent of $P_E(\mathbf{d}_2)$ and $P_E(\mathbf{d}_3)$. From the band structures and the DOS in Supplementary Figs. 16a-c, the shifts in the energy bands and coherence peaks induced by the Doppler shift are clearly evident. Additionally, the simulated real-space LDOS distributions (Supplementary Figs. 16d, e) and their corresponding FFT results (Supplementary Figs. 16f, g) are consistent with the conclusions of our preceding symmetry analysis.

When the magnetic field is applied along the Γ-K direction, the system retains only the $M_y$ mirror symmetry, as can also be seen from the constant energy contours at $E$=0.5Δ in Supplementary Fig. 15. Using the same symmetry analysis, we find that the peak intensities of $\mathbf{d}_2$ and $\mathbf{d}_3$ are equal, while the peak at $\mathbf{d}_1$ is independent. This is consistent with the numerical simulation results shown in Supplementary Fig. 17.

**Supplementary Figures:**

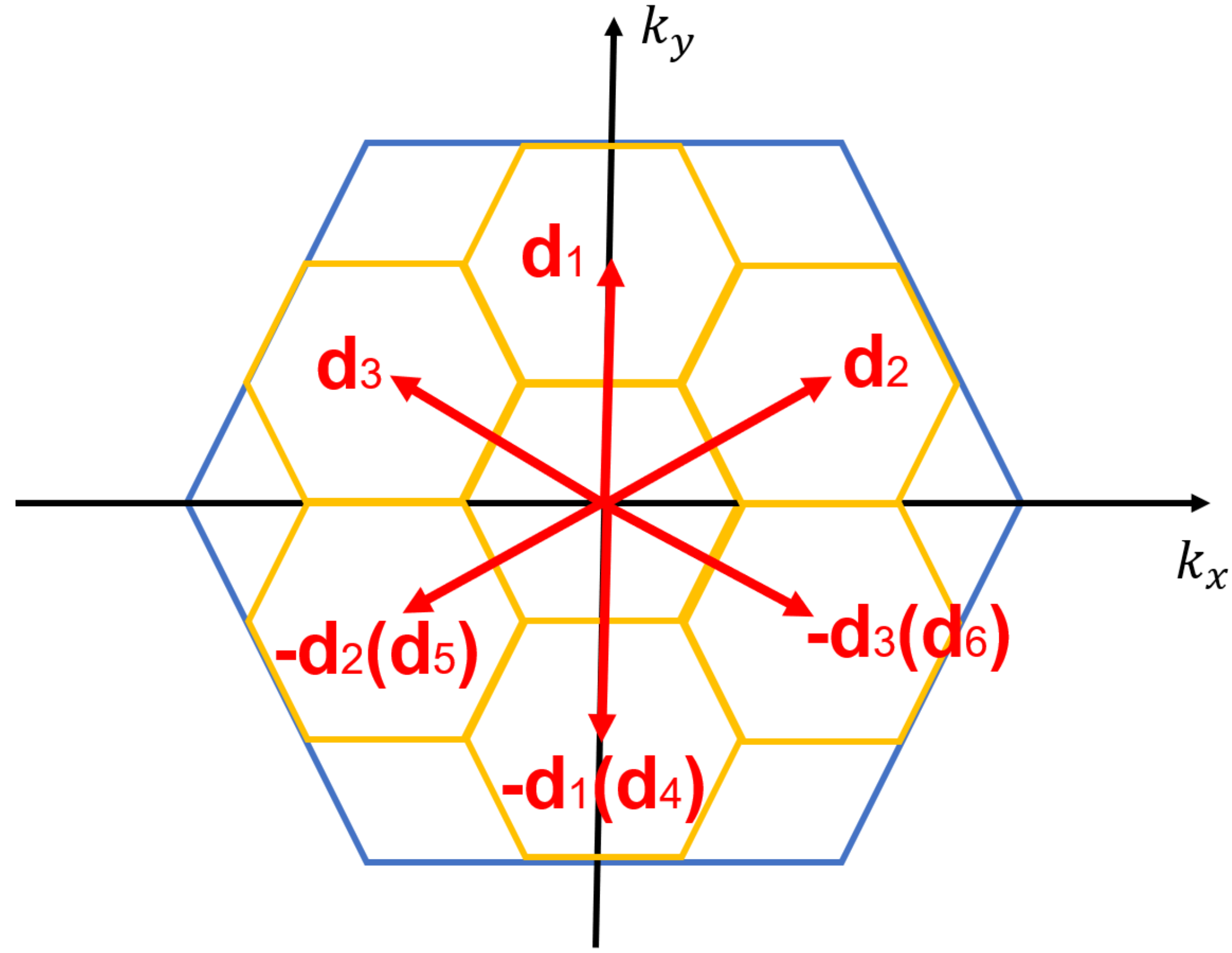

**Supplementary Figure 1| Schematic illustration of the original Brillouin zone and the folded Brillouin zone.** The blue hexagon represents the original Brillouin zone, while the yellow hexagon denotes the folded Brillouin zone after considering the CDW.

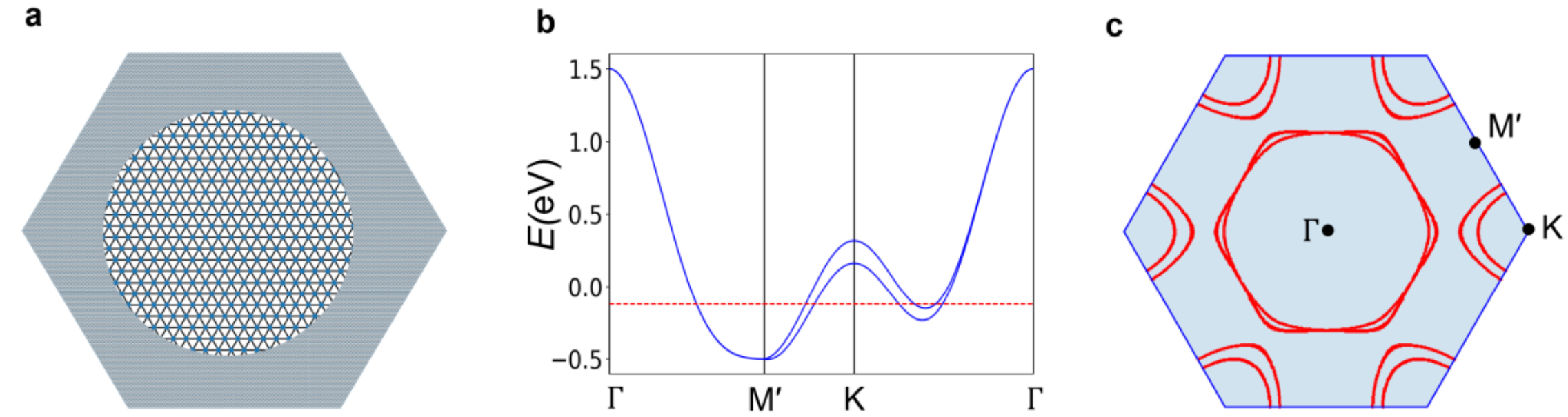

**Supplementary Figure 2|Lattice structure, band structure, and Fermi surface used in the numerical simulations. a,** The surface triangular lattice used for the tight-binding model. **b,** the corresponding tight-binding band structure. **c,** the Fermi surface corresponding to the red line in **b**.

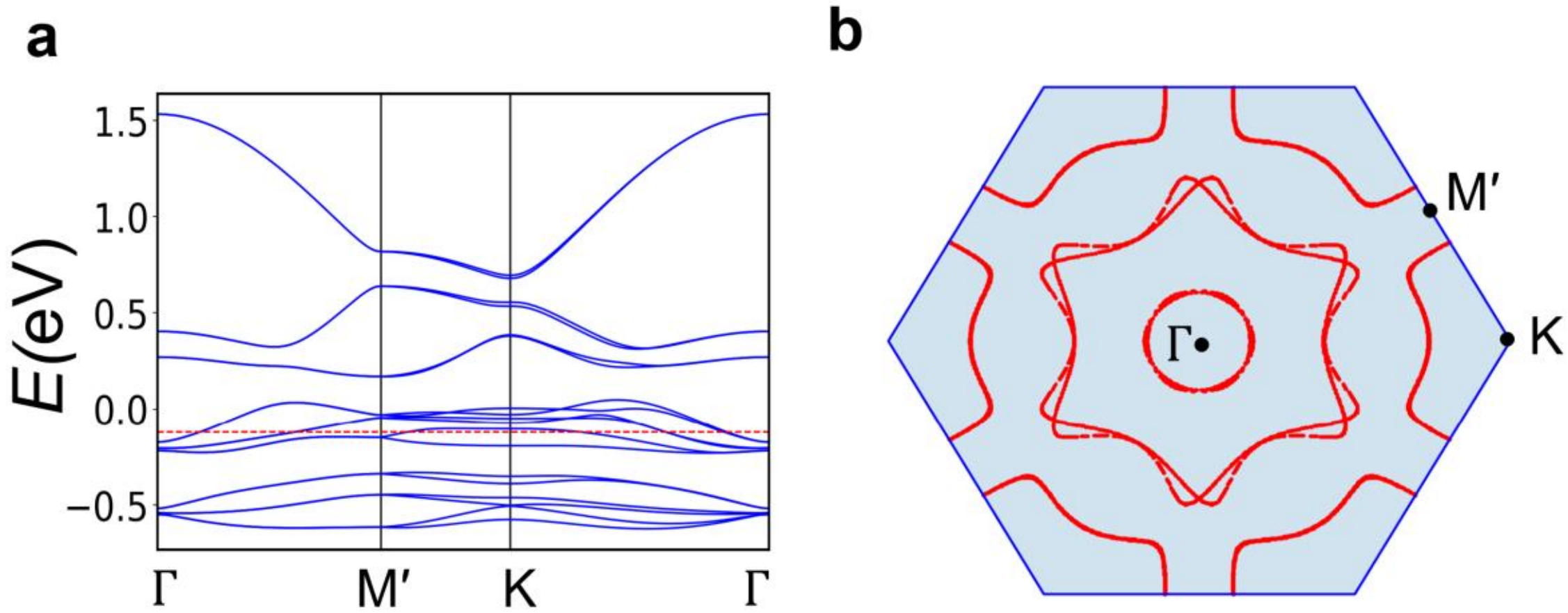

**Supplementary Figure 3| Electronic band structure and Fermi surface after including the CDW.** **a**, Electronic band structure with CDW modulation. **b,** Fermi surface corresponding to the red line in **a**.

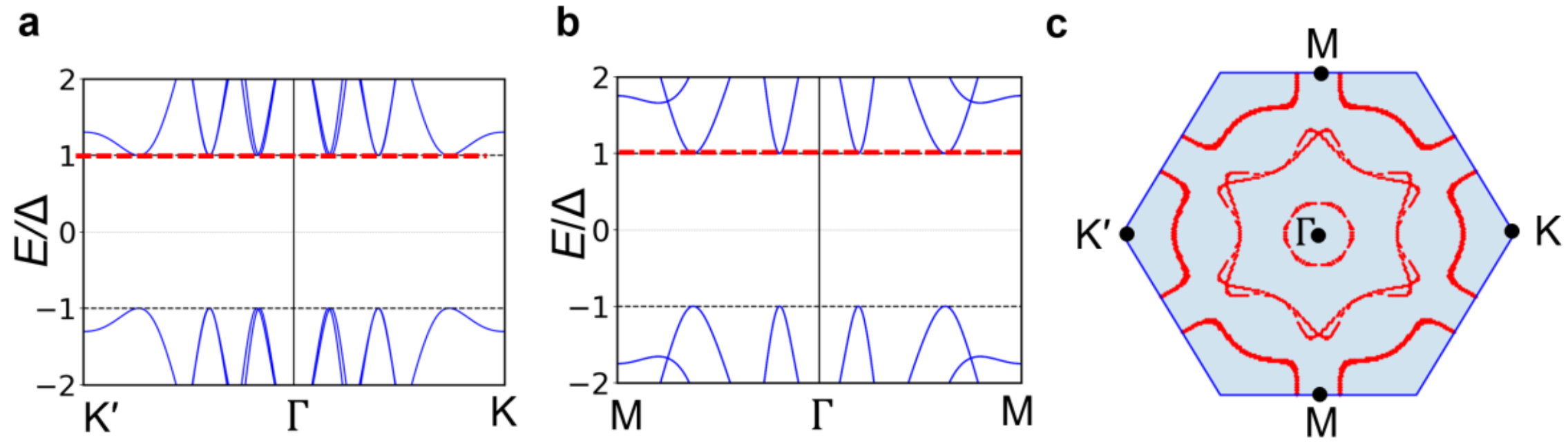

**Supplementary Figure 4| BdG band structure and corresponding constant-energy contour at zero magnetic field. a, b,** The BdG band structure along different high-symmetry paths in the Brillouin zone with zero magnetic field. **c**, Constant-energy contours at $E=\Delta$ corresponding to the red lines in **a, b**.

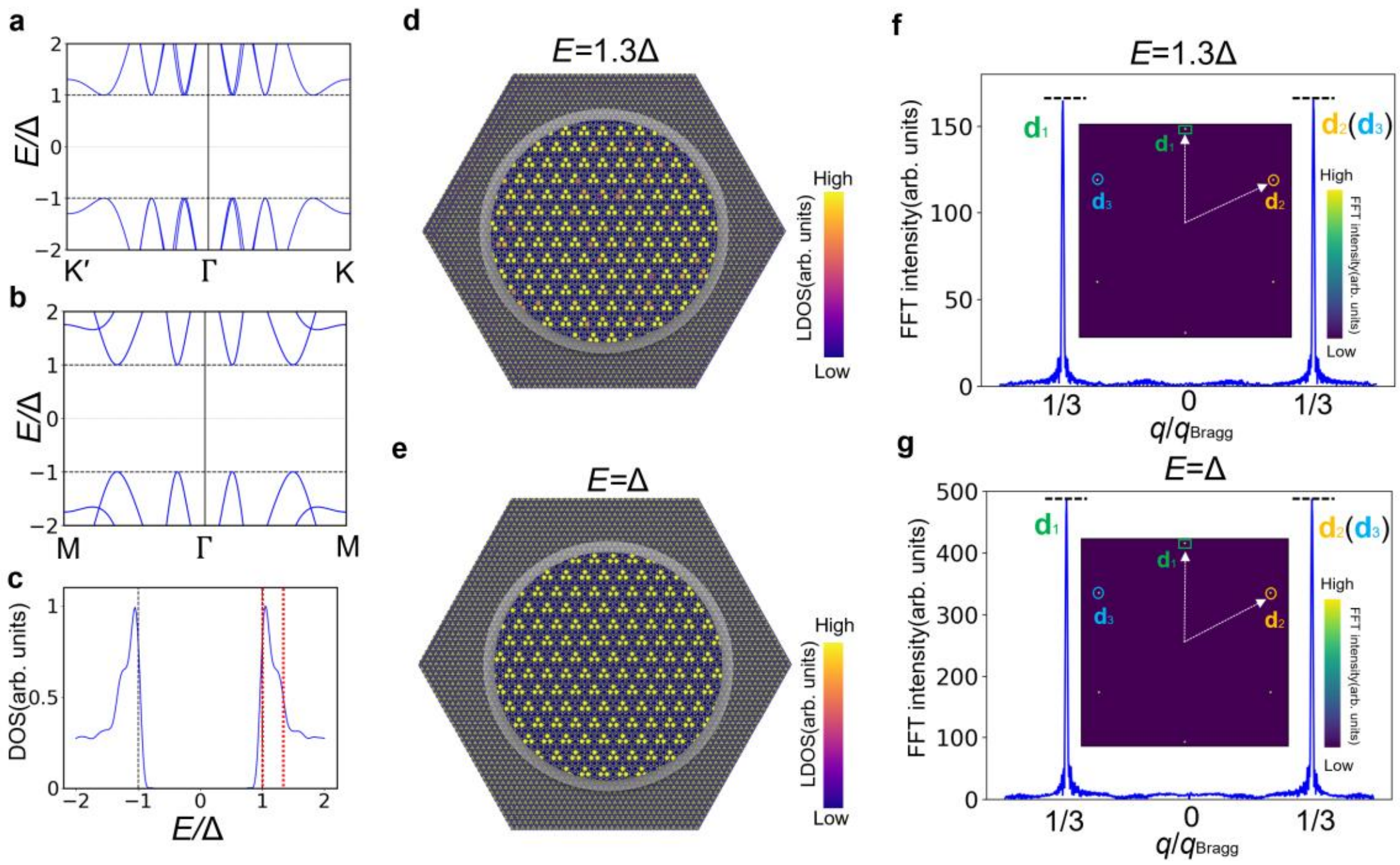

**Supplementary Figure 5| Band structure, DOS, real-space LDOS, and the Fourier transform in the absence of magnetic field. a, b**, The BdG band structure along different high-symmetry paths in the Brillouin zone without an applied magnetic field. **c**, The DOS as a function of energy. **d**, **e**, The real-space distribution of the LDOS after subtracting its spatial average at a given energy **f, g,** Line cuts of the corresponding FFTs (insets) from **d** and **e**, taken along the two directions indicated by dashed arrows.

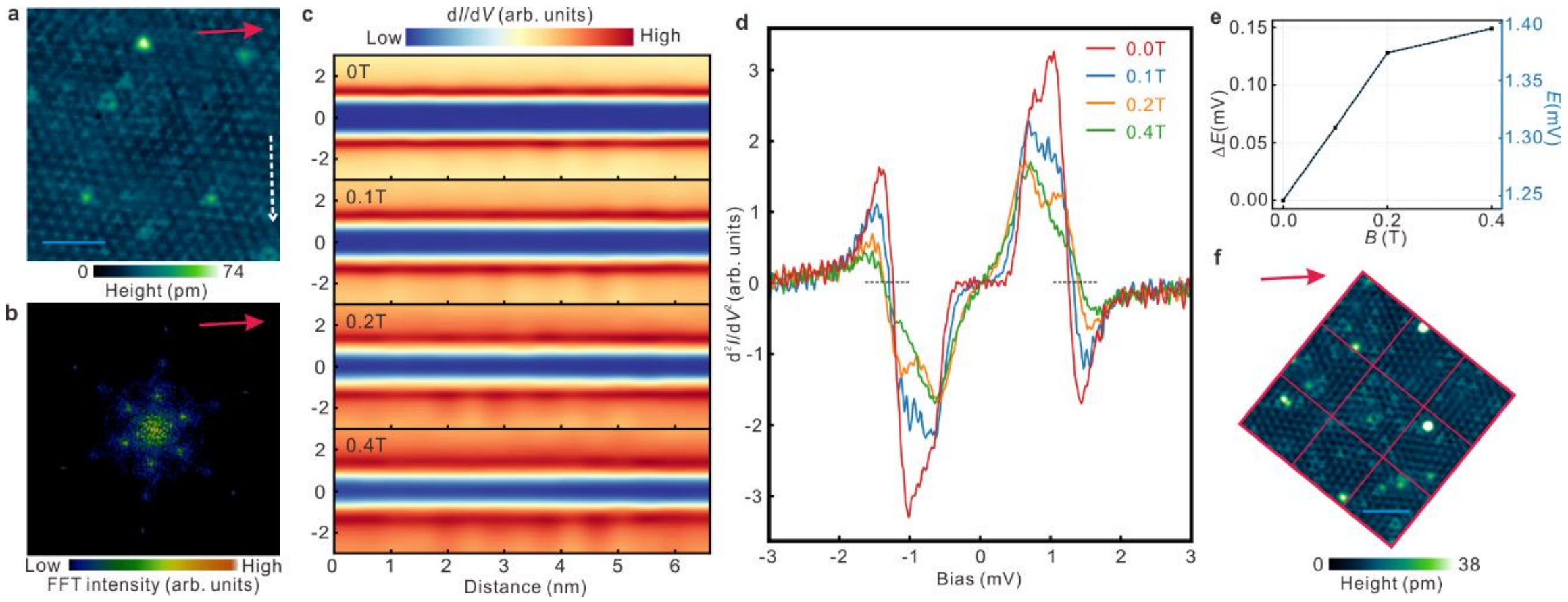

**Supplementary Figure 6| Regions for measurements of field-dependent d*I*/d*V* spectra and the relationship between Doppler shifts and the in-plane magnetic fields**. **a**, Topography of area that measure the field-dependent superconducting gap in Fig.1e ($V$=5 mV, $I$=100 pA). Scale bar, 5nm. The direction of field is indicated by red arrow. **b**, FFT of **a**. **c**, d$I$/d$V$ linecut profile taken under the indicated magnetic field along the white dashed line in **a** ($V$=3 mV, $I$=500 pA, $V_{\mathrm{mod}}$=50 μV). The superconducting gaps shown in Fig. 1e are the average of the corresponding d$I$/d$V$ line cuts measured under each field. Notably, d$I$/d$V$ line cuts across the measured region remain uniform with no appreciable change within the gaps, allowing us to rule out contributions from in-plane vortices. **d**, $d^2I/dV^2$ curves derived from Fig. 1e. **e**, Magnetic-field dependence of the coherence peak energy. **f,** Topography of area that measure the field-dependent spectra in a wide energy range in Fig.1f ($V$=100 mV, $I$=100 pA). The spectra shown in Fig. 1f are the average of spectra measured at the center of each square under corresponding field. Scale bar, 6 nm. The direction of field is indicated by red arrow.

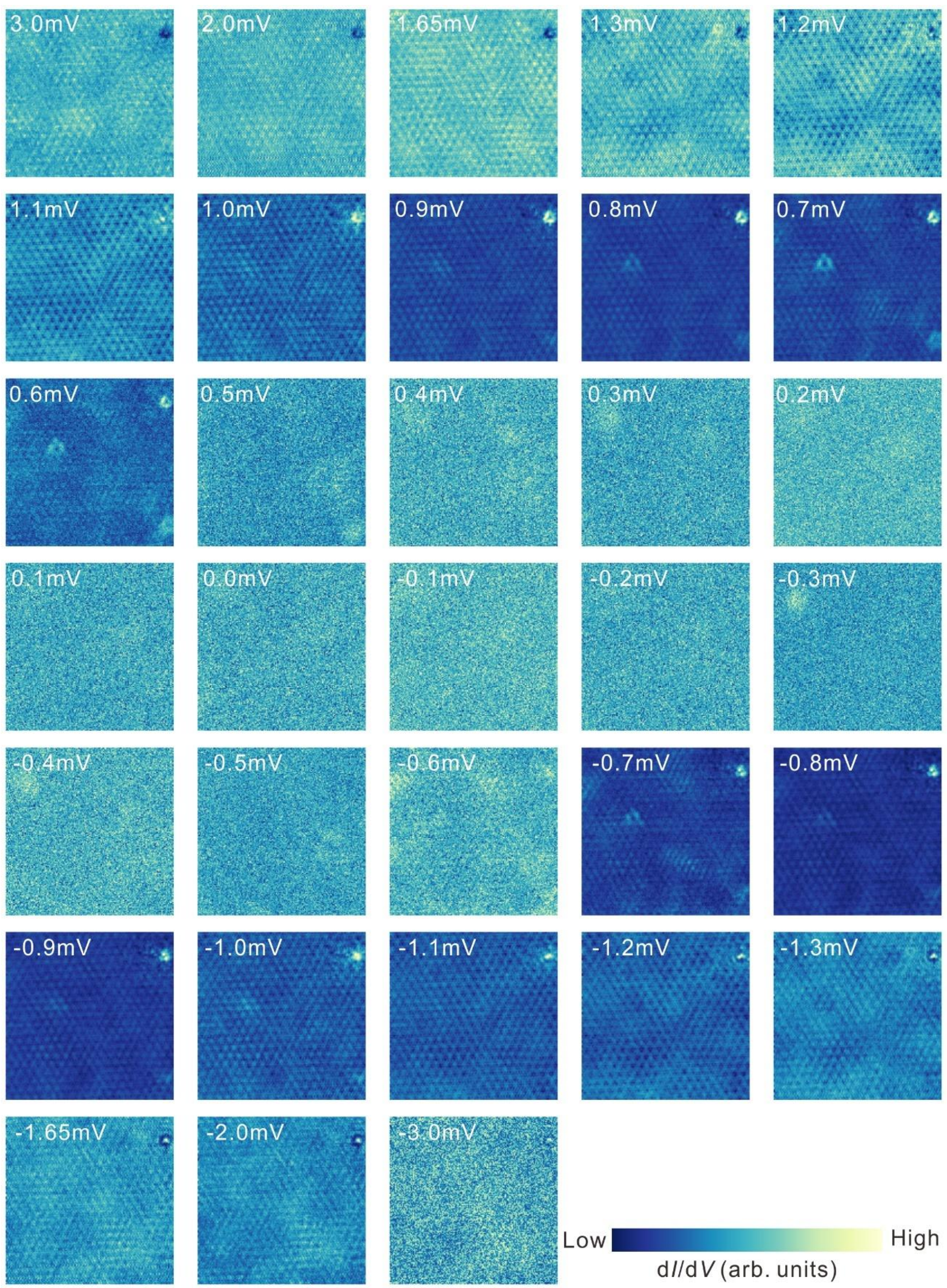

**Supplementary Figure 7| Real-space d*I*/d*V* maps under zero magnetic field**. d*I*/d*V* maps (20×20 nm$^2$) measured at indicated energies (*V*=3 mV, *I*=500 pA). No discernible CDW modulation is visible in the corresponding d*I*/d*V* maps between −0.5 and 0.5 mV, where Bogoliubov quasiparticles are absent.

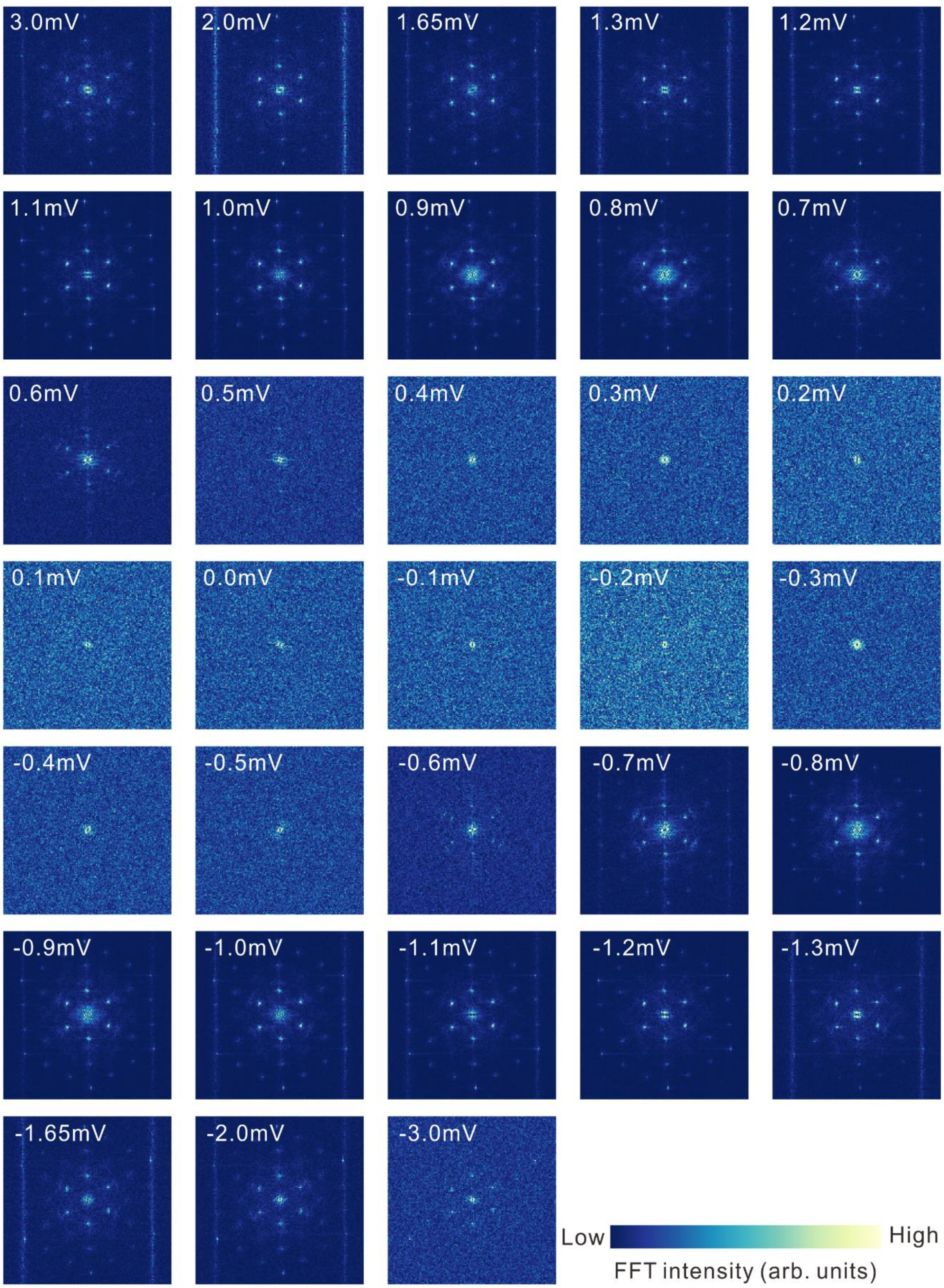

**Supplementary Figure 8| FFTs of real-space d*I*/d*V* maps under zero magnetic field**. FFTs of d*I*/d*V* maps in Supplementary Fig. 7, measured at indicated energies. The CDW signal intensity between −0.5 and 0.5 mV is comparable to the noise level, as no discernible CDW modulation is observed in the corresponding d*I*/d*V* maps.

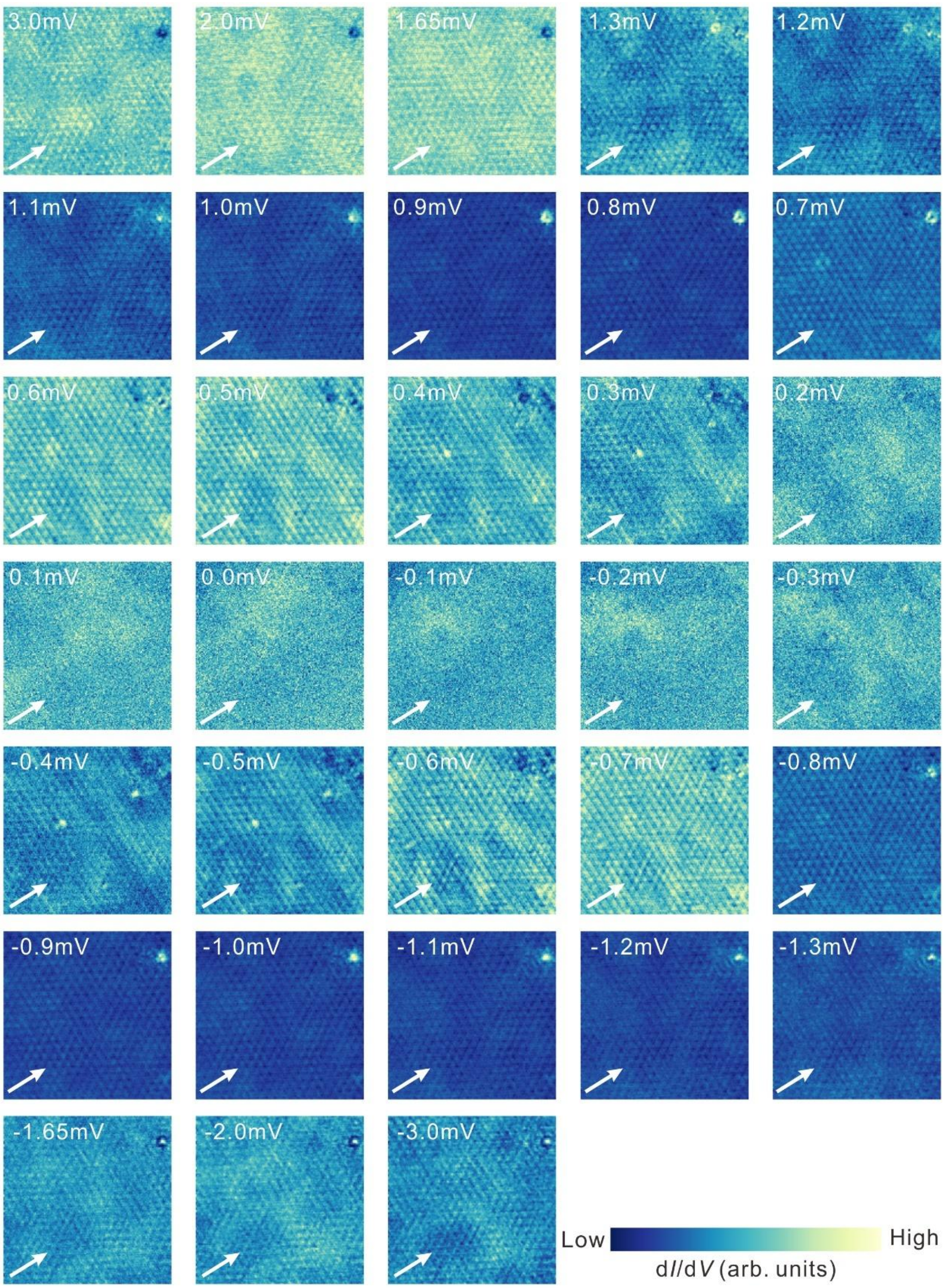

**Supplementary Figure 9| Real-space d*I*/d*V* maps under a 100mT magnetic field applied along Γ-M direction**. d*I*/d*V* maps (20×20 nm$^2$) measured at indicated energies under field applied along Γ-M direction (*V*=3 mV, *I*=500 pA). The measurement area is the same as that shown in Supplementary Fig. 7. No discernible CDW modulation is visible in the corresponding d*I*/d*V* maps between −0.2 and 0.1 mV, where Bogoliubov quasiparticles are absent. The white arrows denote the field directions

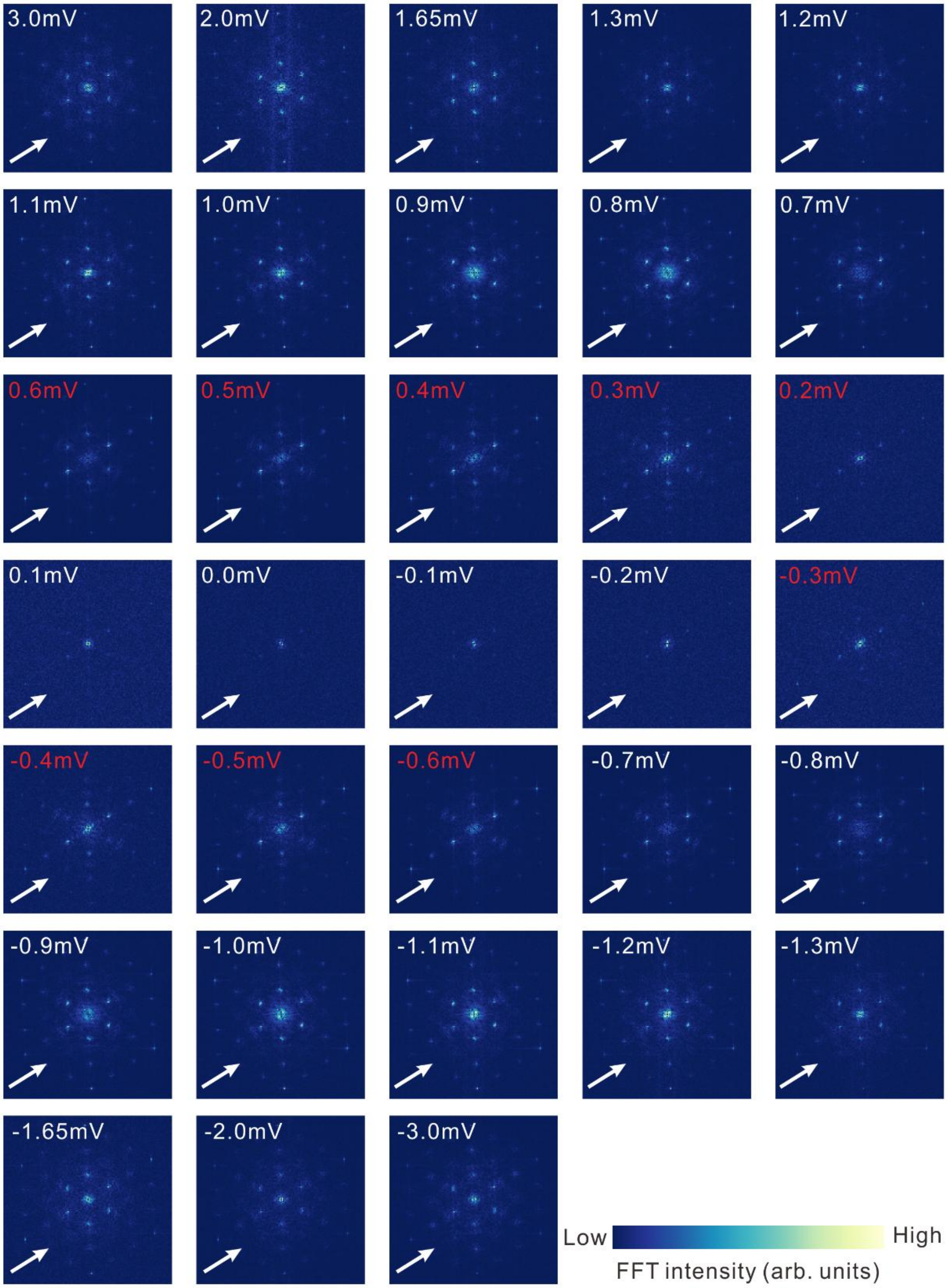

**Supplementary Figure 10| FFTs of real-space d*I*/d*V* maps under a 100mT magnetic field applied along Γ-M direction**. FFTs of d*I*/d*V* maps in Supplementary Fig. 9 measured at indicated energies under field applied along Γ-M direction. Two CDW peaks parallel to the field are selectively enhanced in 0.2 to 0.6 mV and -0.6 to -0.3 mV bias range. The CDW signal intensity between −0.2 and 0.1 mV is comparable to the noise level, as no discernible CDW modulation is observed in the corresponding d*I*/d*V* maps. The white arrows denote the field directions.

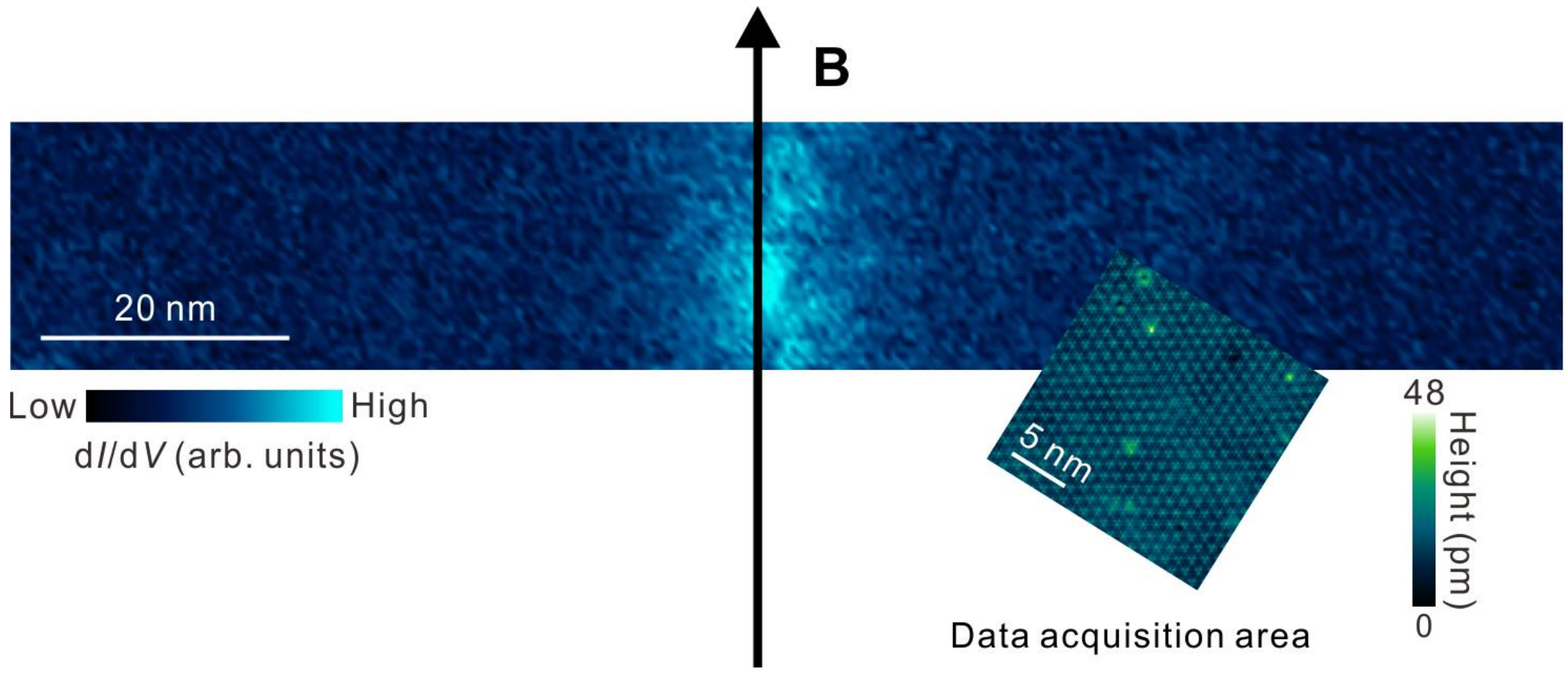

**Supplementary Figure 11| Data acquisition area for Figure 2.** Blue rectangle is the d*I*/d*V* map taken at 0 mV under in-plane field of 100 mT ($V$=3 mV, $I$=200 pA). An in-plane vortex appears in the middle of the map. Green square is the STM image area where the data in Fig. 2 are acquired, away from the in-plane vortex ($V$=3 mV, $I$=100 pA).

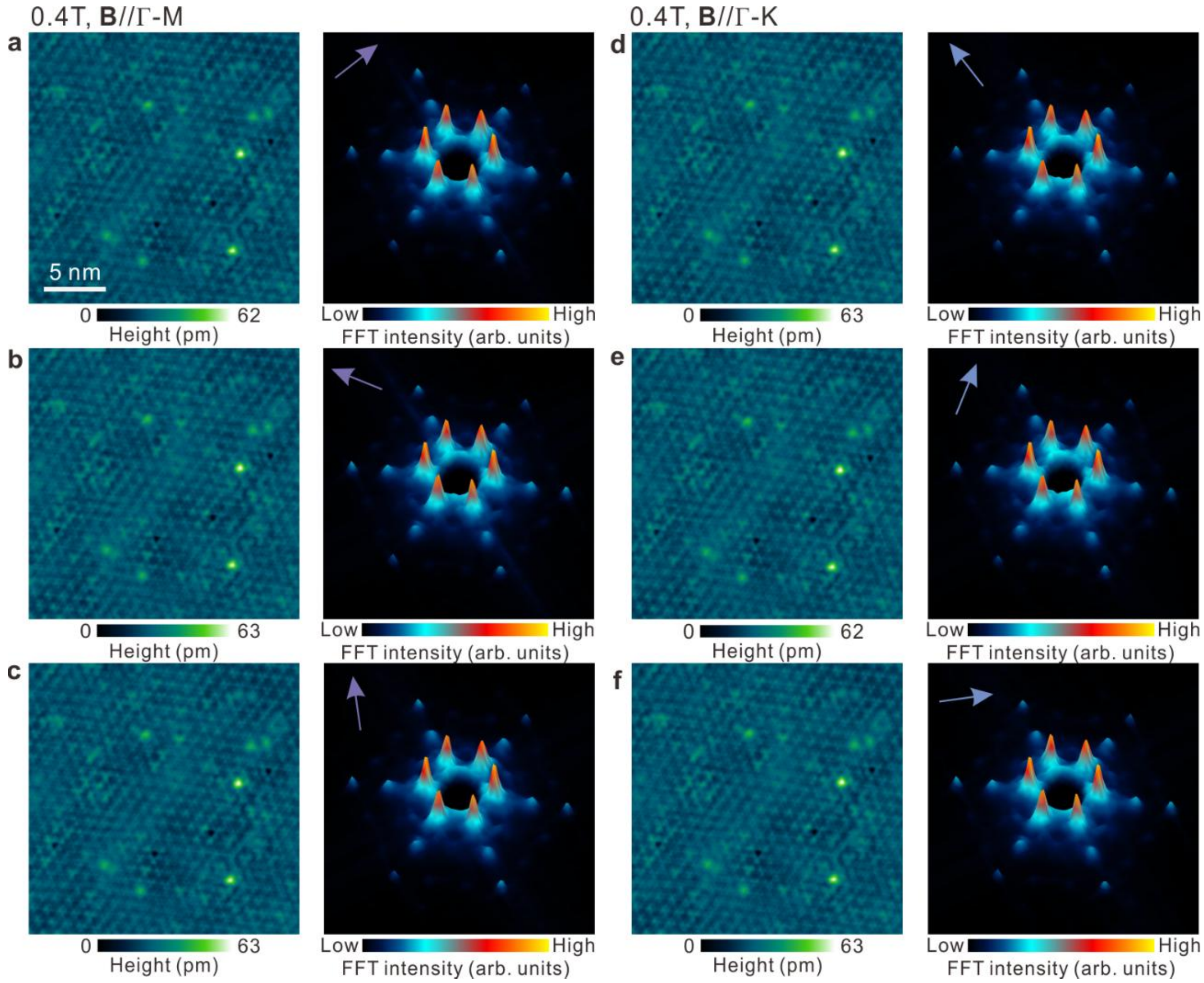

**Supplementary Figure 12| Field-independent CDW modulations in topographic images. a-c**, Topographic images and the corresponding 3D representation of FFTs with magnetic field oriented along the Γ-M directions. **d-f**, Replicating **a-c** with the magnetic field applied along the Γ-K directions. All CDW peaks exhibit comparable intensities and retain $C_{3v}$ symmetry under different magnetic-field orientations. All data were acquired within the same area as Fig. 3 and under an in-plane magnetic field of 400 mT ($V$=5 mV, $I$=0.1 nA). Field orientations are indicated by arrows.

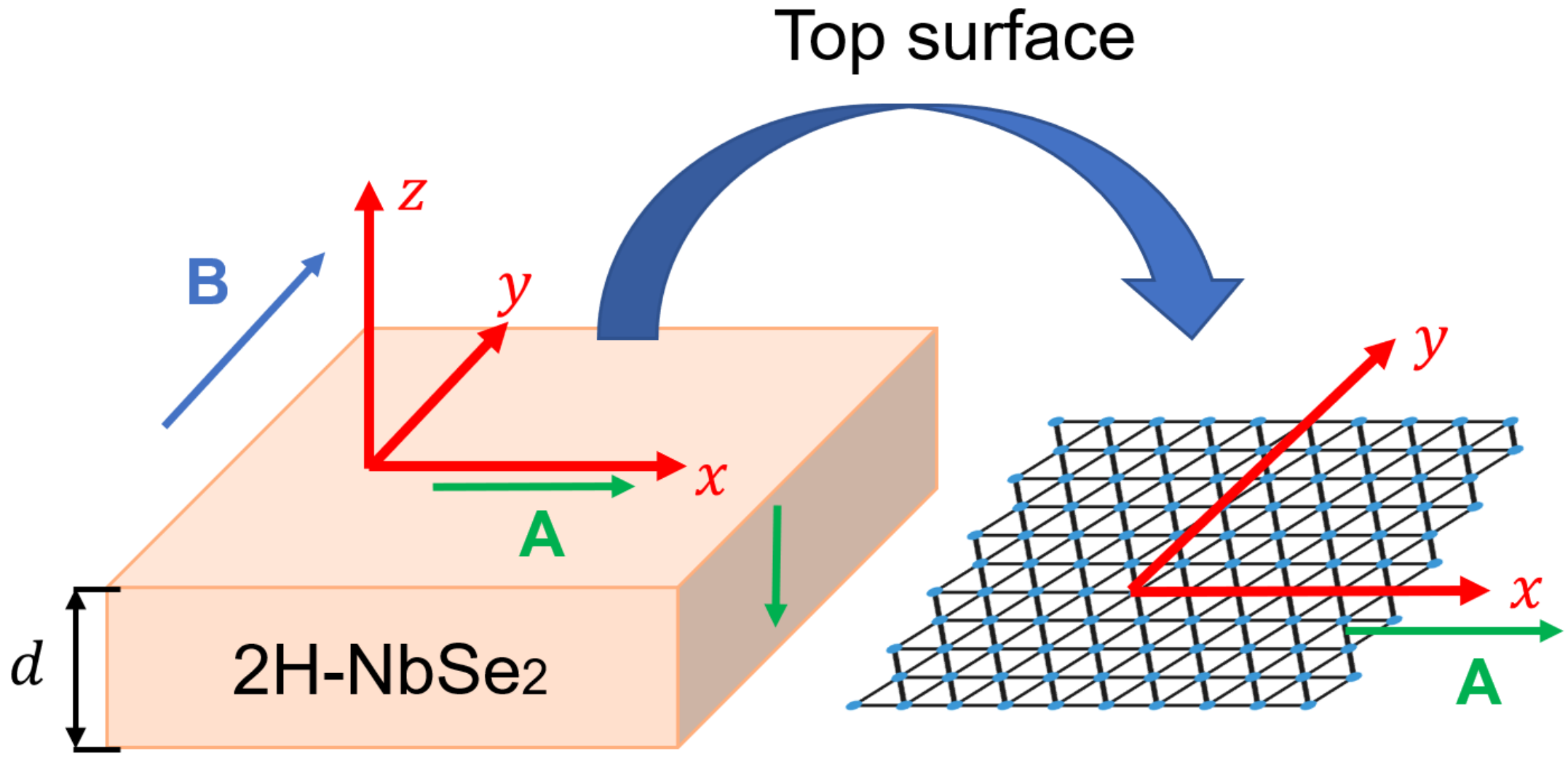

**Supplementary Figure 13| The distribution of the vector potential after applying a magnetic field along the Γ-M direction.** The origin of the coordinates is set at the center, with the top and bottom surfaces located at $z = \pm\frac{d}{2}$.

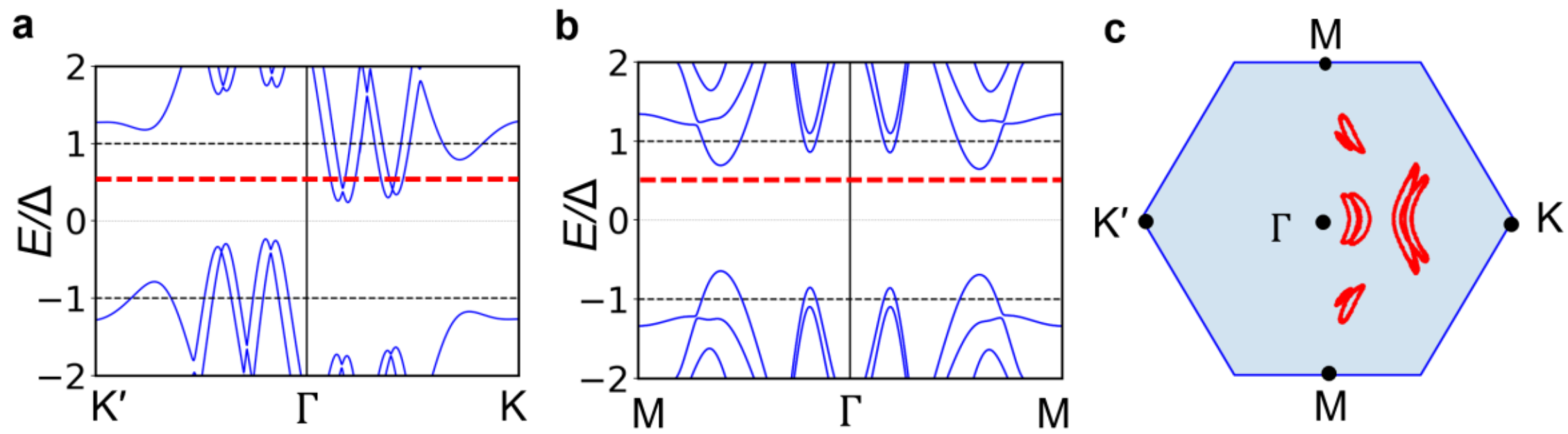

**Supplementary Figure 14| BdG band structure and corresponding constant-energy contour with magnetic field along Γ-M. a, b,** The BdG band structure along different high-symmetry paths in the Brillouin zone with magnetic field along Γ-M. **c**, Constant energy contour at $E$=0.5Δ corresponding to the red lines in **a, b**.

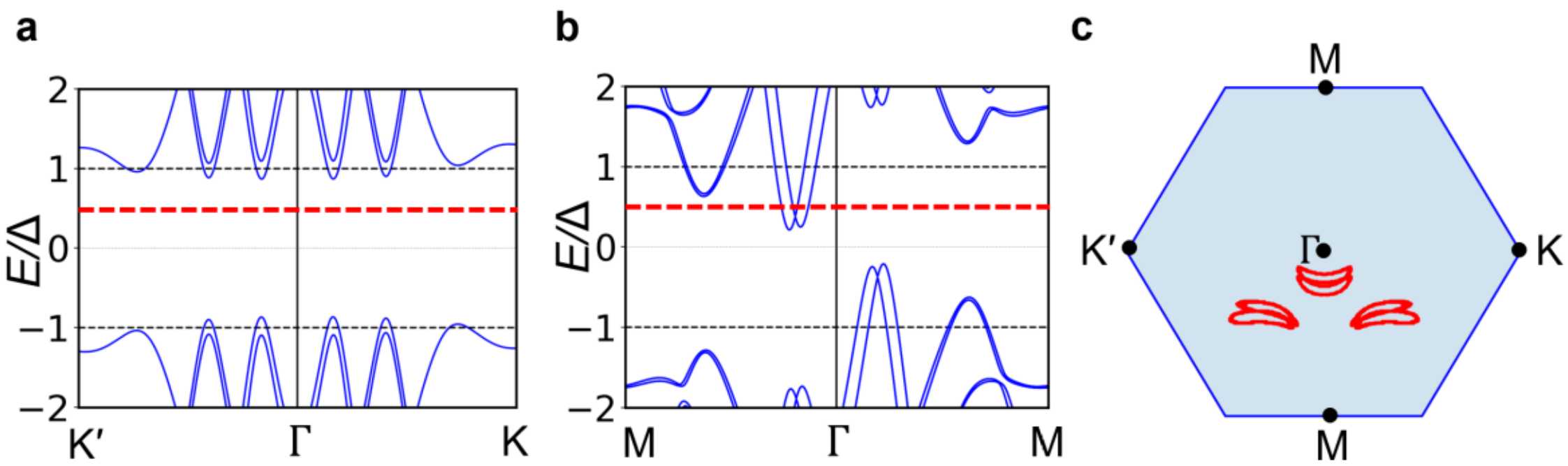

**Supplementary Figure 15| BdG band structure and corresponding constant-energy contour with magnetic field along Γ-K. a, b,** The BdG band structure along different high-symmetry paths in the Brillouin zone with magnetic field along Γ-K. **c**, Constant energy contours at $E$=0.5Δ corresponding to the red lines in **a, b**.

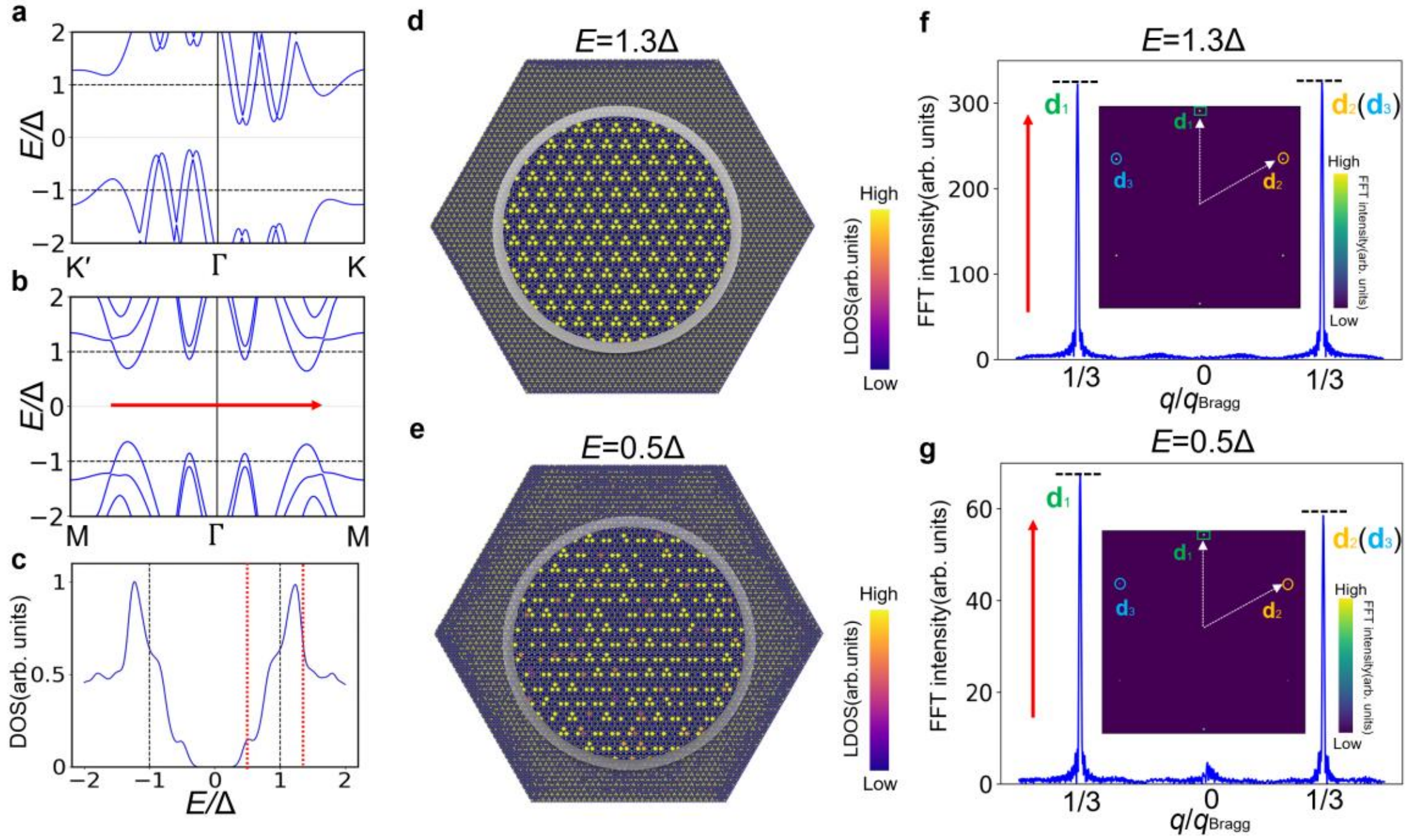

**Supplementary Figure 16| Band structure, DOS, real-space LDOS, and the Fourier transform with magnetic field along Γ-M. a, b,** The BdG band structure along different high-symmetry paths in the Brillouin zone with an applied magnetic field $B$=0.8$B_c$ ($B_c$ is the critical field for closing Δ) along the Γ-M direction. **c,** The DOS as a function of energy. **d, e,** The real-space distribution of the LDOS after subtracting its spatial average at given energies. **f, g,** Line cuts of the corresponding FFTs (insets) from **d** and **e**, taken along the two directions indicated by dashed arrows. All red solid arrows in the figure indicate the direction of the in-plane magnetic field.

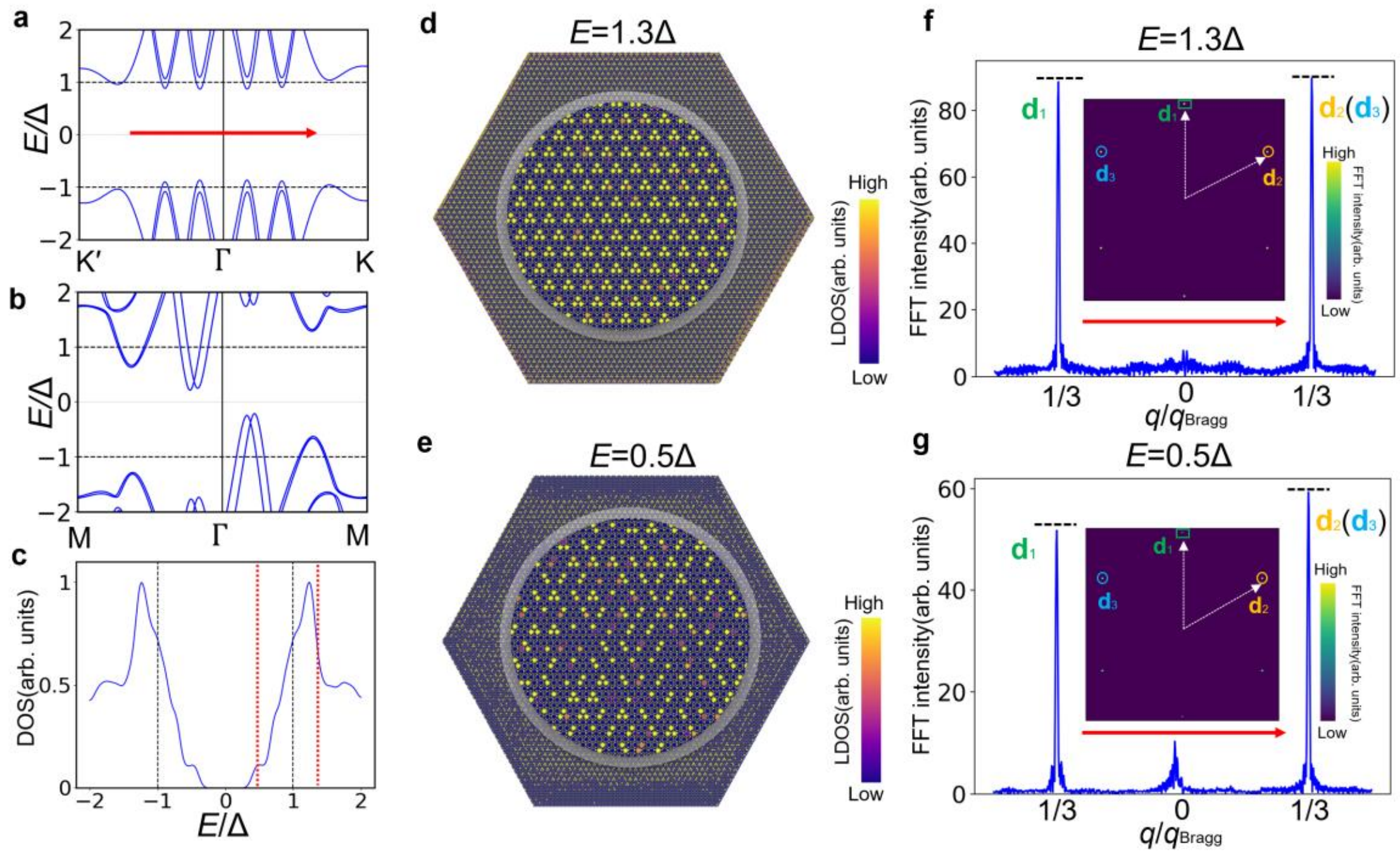

**Supplementary Figure 17| Band structure, DOS, real-space LDOS, and the Fourier transform with magnetic field along Γ-K. a, b,** The BdG band structure along different high-symmetry paths in the Brillouin zone with an applied magnetic field $B$=0.8$B_c$ ($B_c$ is the critical field for closing Δ) along the Γ-K direction. **c,** DOS as a function of energy. **d, e,** The real-space distribution of the LDOS after subtracting its spatial average at given energies. **f, g,** Line cuts of the corresponding FFTs (insets) from **d** and **e**, taken along the two directions indicated by dashed arrows. All red solid arrows in the figure indicate the direction of the in-plane magnetic field.